\documentclass[12pt]{article}
\usepackage[dvips]{graphicx}
\usepackage{rotating}

\usepackage{epsfig,amsmath,amssymb,verbatim,mathrsfs}

\numberwithin{equation}{section}

\def\beq{\begin{eqnarray}}
\def\eeq{\end{eqnarray}}
\def\bea{\begin{eqnarray}}
\def\eea{\end{eqnarray}}

\def\tev{\, {\rm TeV}}
\def\gev{\, {\rm GeV}}
\def\mev{\, {\rm MeV}}
\def\kev{\, {\rm keV}}
\def\kevee{\, {\rm keVee}}

\newcommand{\gsim}{\lower.7ex\hbox{$\;\stackrel{\textstyle>}{\sim}\;$}}
\newcommand{\lsim}{\lower.7ex\hbox{$\;\stackrel{\textstyle<}{\sim}\;$}}

\def\stilde{\widetilde}

\newcommand{\newc}{\newcommand}
\newc{\Nc}{N_{c}}
\newc{\CG}{C_G}
\newc{\gp}{g'}
\newc{\stopi}{\stilde t_i}
\newc{\sboti}{\stilde b_i}
\newc{\staui}{\stilde \tau_i}
\newc{\stopj}{\stilde t_j}
\newc{\sbotj}{\stilde b_j}
\newc{\stauj}{\stilde \tau_j}
\newc{\stopI}{\stilde t_1}
\newc{\stopII}{\stilde t_2}
\newc{\sbotI}{\stilde b_1}
\newc{\sbotII}{\stilde b_2}
\newc{\stauI}{\stilde \tau_1}
\newc{\stauII}{\stilde \tau_2}
\newc{\sstop}{s_{t}}
\newc{\cstop}{c_{t}}
\newc{\ssbot}{s_{b}}
\newc{\csbot}{c_{b}}
\newc{\sstau}{s_{\tau}}
\newc{\cstau}{c_{\tau}}
\newc{\Sstop}{s_{2t}}
\newc{\Cstop}{c_{2t}}
\newc{\Ssbot}{s_{2b}}
\newc{\Csbot}{c_{2b}}
\newc{\Sstau}{s_{2\tau}}
\newc{\Cstau}{c_{2\tau}}
\newc{\salpha}{s_\alpha}
\newc{\calpha}{c_\alpha}
\newc{\Calpha}{c_{2\alpha}}
\newc{\Salpha}{s_{2\alpha}}
\newc{\sbetapm}{s_{\beta_\pm}}
\newc{\cbetapm}{c_{\beta_\pm}}
\newc{\Sbetapm}{s_{2 \beta_\pm}}
\newc{\Cbetapm}{c_{2 \beta_\pm}}
\newc{\sbetaO}{s_{\beta_0}}
\newc{\cbetaO}{c_{\beta_0}}
\newc{\SbetaO}{s_{2 \beta_0}}
\newc{\CbetaO}{c_{2 \beta_0}}
\newc{\vu}{v_u}
\newc{\vd}{v_d}
\newc{\seL}{\stilde e_L}
\newc{\smuL}{\stilde \mu_L}
\newc{\seR}{\stilde e_R}
\newc{\smuR}{\stilde \mu_R}
\newc{\suL}{\stilde u_L}
\newc{\sdL}{\stilde d_L}
\newc{\suR}{\stilde u_R}
\newc{\sdR}{\stilde d_R}
\newc{\scL}{\stilde c_L}
\newc{\ssL}{\stilde s_L}
\newc{\scR}{\stilde c_R}
\newc{\ssR}{\stilde s_R}
\newc{\snue}{\stilde \nu_e}
\newc{\snumu}{\stilde \nu_\mu}
\newc{\snutau}{\stilde \nu_\tau}
\newc{\Gpm}{G^\pm}
\newc{\Hpm}{H^\pm}
\newc{\FFbS}{\overline{FF}S}
\newc{\FFbV}{\overline{FF}V}
\newc{\FSS}{F_{SS}}
\newc{\FSSS}{F_{SSS}}
\newc{\FFFS}{F_{FFS}}
\newc{\FFFbS}{F_{\overline{FF}S}}
\newc{\FSSV}{F_{SSV}}
\newc{\FVS}{F_{VS}}
\newc{\FVVS}{F_{VVS}}
\newc{\FFFV}{F_{FFV}}
\newc{\FFFbV}{F_{\overline{FF}V}}
\newc{\Fgauge}{F_{\rm gauge}}
\newc{\DRbarprime}{$\overline{\rm DR}'$ }
\newc{\DRbar}{$\overline{\rm DR}$ }
\newc{\MSbar}{$\overline{\rm MS}$ }
\newc{\Yu}{{\bf Y}_u}
\newc{\Yd}{{\bf Y}_d}
\newc{\Ye}{{\bf Y}_e}
\newc{\Au}{{\bf a}_u}
\newc{\Ad}{{\bf a}_d}
\newc{\Ae}{{\bf a}_e}
\newc{\bm}{{\bf m}}
\newc{\zhol}{Z^{\rm hol}}

\newcommand{\ccdot}{\!\cdot\!}
\newcommand{\nnmb}{\nonumber}
\newcommand{\del}{\partial}

\newcommand{\lrf}[2]{\left(\frac{#1}{#2}\right)}

\begin{document}

\setlength{\baselineskip}{0.2in}



\begin{titlepage}
\noindent
\begin{flushright}
\end{flushright}
\vspace{1cm}

\begin{center}
  \begin{Large}
    \begin{bf}
Candidates for Inelastic Dark Matter\\
     \end{bf}
  \end{Large}
\end{center}
\vspace{0.2cm}

\begin{center}

\begin{large}
Yanou Cui, David E. Morrissey, David Poland, and Lisa Randall\\
\end{large}
\vspace{0.3cm}
  \begin{it}
Jefferson Physical Laboratory, Harvard University,\\
Cambridge, Massachusetts 02138, USA
\vspace{0.5cm}
\end{it}\\

\end{center}

\center{\today}

\begin{abstract}

  Although we have yet to determine whether the DAMA data
represents a true discovery of new physics, among such
interpretations inelastic dark matter~(IDM) can match the energy
spectrum of DAMA very well while not contradicting the results of
other direct detection searches. In this paper we investigate the
general properties that a viable IDM candidate must have and search
for simple models that realize these properties in natural ways.  We
begin by  determining the regions of IDM parameter space that are
allowed by direct detection searches including DAMA, paying special
attention to larger IDM masses.  We observe that an inelastic dark
matter candidate  with electroweak interactions can naturally
satisfy observational constraints while simultaneously yielding the
correct thermal relic abundance.  We comment on several other
proposed dark matter explanations for the DAMA signal and
demonstrate that one of the proposed alternatives -- elastic
scattering of dark matter off electrons -- is strongly disfavored
when the modulated and unmodulated DAMA spectral data are taken into
account.  We then outline the general essential features of IDM
models in which inelastic scattering off nuclei is mediated by the
exchange of a massive gauge boson, and construct natural models in
the context of a warped extra dimension and supersymmetry.

\end{abstract}

\vspace{1cm}

\end{titlepage}

\setcounter{page}{2}

\tableofcontents

\vfill\eject



\newpage

\section{Introduction}

The DAMA/NaI and DAMA/LIBRA experiments observe
an annual modulation signal
in their NaI-based scintillation detectors
with a statistical significance of $8.3\,\sigma$~\cite{Bernabei:2008yi}.
A possible origin of this signal is galactic dark matter~(DM)
scattering off the nuclei in the detectors of these
experiments~\cite{Goodman:1984dc,Drukier:1986tm}.
The annual variation would then result from the motion of the Earth
relative to the Sun as it passes through the halo of dark matter
enveloping our galaxy.  The phase, period, and  amplitude
of the modulation signal seen by DAMA are all consistent with
DM scattering~\cite{Bernabei:2008yi}. We have yet to determine whether
the observation is the result of truly new physics, or if it results
from an unaccounted-for detector effect or background. In this paper,
we ask the question: what if the DAMA result is truly a discovery
of dark matter?  What could it possibly be?

  The major challenge for a DM interpretation of the DAMA result is
that it appears to be at odds with the bounds on coherent DM-nucleus
scattering obtained by other DM direct detection experiments
such as CDMS~\cite{Ahmed:2008eu} and XENON~\cite{Angle:2007uj}.
In contrast to DAMA, these experiments search for an unmodulated nuclear
recoil signal from DM scattering using stronger background rejection
methods.  The bounds they place on DM-nucleon cross sections
rule out coherent elastic scattering off iodine nuclei as the origin
of the DAMA signal by several orders of magnitude.  Lighter DM scattering
coherently off sodium nuclei is marginally consistent with both DAMA and
other experiments~\cite{Gelmini:2004gm,Petriello:2008jj},
but gives a very poor fit to the energy spectrum of the
modulated DAMA signal~\cite{Chang:2008xa,Fairbairn:2008gz,Savage:2008er}.

  An elegant possibility that can account for the signal observed
by the DAMA experiments that is also consistent with other direct detection
experiments is dark matter that scatters \emph{inelastically} off
nuclei~\cite{TuckerSmith:2001hy,TuckerSmith:2002af,
TuckerSmith:2004jv,Chang:2008gd}.  In the inelastic dark matter~(IDM)
scenario, the dark matter particle $\chi_1$ scatters preferentially
off target nuclei into a slightly heavier $\chi_2$ state.
The kinematics of this process can enhance the nuclear recoil signal
at DAMA relative to other experiments such as CDMS in a couple of ways.
To produce a nuclear recoil signal with energy $E_R$, the minimum
incident velocity of the DM particle is~\cite{TuckerSmith:2001hy}
\beq
v_{min} = \frac{1}{\sqrt{2\,m_N\,E_R}}\,
\left(\frac{m_N\,E_R}{\mu_N}+\delta\right),
\label{vmin}
\eeq
where $\delta$ is the mass splitting between $\chi_1$ and $\chi_2$,
$m_N$ is the mass of the target nucleus, and $\mu_N$ is the reduced
mass of the nucleus-DM system.  The distribution
of DM velocities in the galactic halo is expected
to be approximately Maxwellian with an upper cutoff at the galactic escape
velocity $v_{esc}$~\cite{Drukier:1986tm}.  When the second term in this
expression dominates, the minimal velocity needed to produce
a recoil energy $E_R$ is lower for heavier nuclei.  This leads to an
enhanced signal at DAMA, which contains iodine with $A\simeq 127$
as a detector material, relative to CDMS, consisting of germanium
with $A\simeq 73$~\cite{TuckerSmith:2001hy}.  The kinematics of
inelastic DM scattering also increases the amount of annual modulation
compared to the unmodulated signal rate, further enhancing the signal
at DAMA relative to other direct detection
experiments~\cite{TuckerSmith:2001hy}.

  While IDM provides a compelling explanation for the DAMA signal,
only a few concrete particle physics candidates have been
proposed. A model with sneutrino DM where the inelastic splitting
is induced by the lepton-number violating superpotential operator
$W\supset (L\ccdot H_u)^2/\Lambda$ was suggested in
Ref.~\cite{TuckerSmith:2001hy}.  Ref.~\cite{ArkaniHamed:2008qn}
proposed that the inelasticity could arise from the radiative
splitting of masses within a multiplet after the spontaneous
breakdown of a new non-Abelian hidden gauge symmetry around a GeV.
Pseudo-Dirac neutralinos as IDM arising from from approximately
$R$-symmetric SUSY scenarios are considered in
Refs.~\cite{rsymidm1,rsymidm2}.

  In the present work we seek to obtain a broader overview of
potential candidates for IDM.  We study the general features
required for IDM to account for the DAMA signal and we describe
several explicit IDM candidates.  To remain as general as possible,
we do not attempt to account for the tantalizing indirect hints
for dark matter such as the excess positron and electron fluxes
observed by PAMELA~\cite{Adriani:2008zr}, ATIC~\cite{Chang:2008zz},
and PPB-BETS~\cite{Torii:2008xu},
the INTEGRAL $511\,\kev$ line~\cite{Weidenspointner:2006nua},
or the WMAP haze~\cite{Finkbeiner:2003im,Dobler:2007wv,Hooper:2007kb}.
  In  this paper our goal is simply to understand the possibilities for DAMA alone.
What if DAMA represents a true discovery?
What would be credible candidates for what it could be?
It would of course be interesting to determine which of the
candidates for IDM might be compatible with these indirect signals,
but we postpone this direction to future work.

  The outline of this paper is as follows.  In Section~\ref{fits}
we investigate how well the hypothesis of IDM can account for the
DAMA signal while evading the constraints from other direct
detection searches for dark matter.  We also argue that several
other mechanisms proposed to explain this puzzle do not give a
good fit to the full DAMA dataset. In Section~\ref{general} we
discuss the general features required for a model to generate IDM,
and we study the corresponding phenomenological constraints on
these features.  We present several plausible and explicit models
that can give rise to IDM in Section~\ref{models}.
Section~\ref{concl} is reserved for our conclusions.

\section{Inelastic Dark Matter as an Explanation for DAMA\label{fits}}

  Inelastic dark matter has been shown to provide a compelling
explanation for the signal in the DAMA experiments while remaining
compatible with other direct DM searches such as CDMS, XENON, and
CRESST for dark matter masses at least as large as
$250\,\gev$~\cite{Chang:2008gd}.  Here we extend the analysis of
Ref.~\cite{Chang:2008gd} to larger dark matter masses. Such larger masses are of interest
theoretically because they can be natural in the context of models addressing
the hierarchy problem such as warped geometry~\cite{Ponton:2008zv}
and supersymmetry~\cite{ArkaniHamed:2006mb}. In the case of
electroweakly charged dark matter, they also lead to the correct
thermal relic abundance.  Additional attention to heavy dark matter
is due to the recent experimental results from ATIC and PPB-BETS.
We also consider some of the proposed alternative
explanations of the DAMA signal and comment on their viability.

\subsection{IDM Fits to the DAMA Data}

  We begin by reviewing the formalism for calculating the
expected signal at direct detection experiments from the scattering
of dark matter. The total rate of inelastic nuclear recoil
scatterings per unit mass of detector per unit recoil energy $E_R$
in the lab frame is
\beq
\frac{dR}{dE_R} =
N_T\frac{\rho_{DM}}{M_{DM}}\, \int_{v_{min}}\!d^3v\,
v\,f(\vec{v},\vec{v}_e)\,\frac{d\sigma}{dE_R},
\label{rateeq}
\eeq
where $\rho_{DM} \simeq 0.3\,\gev/cm^3$ is the local DM density,
$M_{DM}$ is the DM mass, $N_T$ is the number of target nuclei per
unit mass of detector,
and $f(\vec{v},\vec{v}_e)$ is the local
dark matter velocity distribution.  For coherent spin-independent DM
scattering, the DM-nucleus differential cross section $d\sigma/dE_R$
has the general form~\cite{Jungman:1995df,Bertone:2004pz}
\beq
\frac{d\sigma}{dE_R}
= \frac{1}{v^2}\frac{m_N\,\sigma_n^0}{2\mu_n^2}\,
\frac{\left[f_p\,Z+f_n\,(A-Z)\right]^2}{f_n^2}\,F^2(E_R),
\label{dsigder}
\eeq
where $m_N$ is the mass of the target nucleus
with atomic and molecular numbers $A$ and $Z$, $\mu_n$ is the
DM-nucleon reduced mass,
$f_p$ and $f_n$ are effective coherent couplings to the proton and
neutron, and $\sigma_n^0$ is the overall effective DM-neutron
cross section at zero momentum transfer. The function $F^2(E_R)$
is a form factor characterizing the
loss of coherence as the momentum transfer $q^2 = 2\,m_N\,E_R$
deviates from zero. For computational simplicity, we use the
Helm/Lewin-Smith~\cite{Helm:1956zz,Lewin:1995rx} parameterization of
the form factor.  In order to correct for the fact that this
parameterization can be off by $\sim 20 \%$ for larger values of
$E_R$, we weight this by a quartic polynomial fit to the table given
in~\cite{Duda:2006uk}, which gives the ratios of the
Helm/Lewin-Smith form factor to the more accurate Two-Parameter
Fermi (Woods-Saxon) form factor for various elements and values of
$E_R$.

   We take the DM velocity distribution to be Maxwellian with
a cutoff~\cite{Drukier:1986tm},
\beq f(\vec{v},\vec{v}_e) =
\frac{1}{(\pi\,v_0^2)^{3/2}}\,
e^{-(\vec{v}+\vec{v}_e)^2/v_0^2}\;
\Theta(v_{esc}-\left|\vec{v}+\vec{v}_e\right|).
\eeq
Here, $v_0 \simeq 220\,km/s$ is the DM \emph{rms} speed, $v_{esc}$ is
the local DM \emph{escape velocity} in the halo frame, $\vec{v}_e$
is the velocity of the Earth with respect to the galactic DM halo,
and $\vec{v}$ is the DM velocity in the Earth frame. The signals
from IDM are extremely sensitive to the value of $v_{esc}$, which is
thought to be in the range $498\,km/s < v_{esc} <
608\,km/s$~\cite{Smith:2006ym}. The velocity of the Earth relative to the halo,
$\vec{v}_e$, has components from both the motion of the solar system
relative to the halo as well as the annual motion of the Earth about
the Sun,
\beq
\vec{v}_e(t) = \vec{v}_s + V_o \left[ \hat{\epsilon}_1
\cos(2 \pi (t - t_1)) + \hat{\epsilon}_2 \sin(2 \pi (t - t_1))
\right]
\eeq
where $\vec{v}_s \simeq (0,220,0) + (10,5,7) ~km/s$ is
the velocity of the solar system relative to the
halo~\cite{Dehnen:1997cq,Binney:2008},
$V_o \simeq 29.79 ~km/s$ is the Earth's orbital speed~\cite{Lewin:1995rx},
and $t$ is measured in years. Following the conventions
of Ref.~\cite{Gelmini:2000dm} and the discussion in Ref.~\cite{Savage:2008er},
we are using coordinates where $\hat{x}$ points to the center of the galaxy,
$\hat{y}$ gives the direction of disk rotation, and $\hat{z}$ points to the
north galactic pole. The directions of the Earth's motion on
$t_1 = \text{March 21}$
($\hat{\epsilon}_1$) and June 21 ($\hat{\epsilon}_2$) are given by
$\hat{\epsilon}_1 = (0.9931, 0.1170, -0.01032)$ and
$\hat{\epsilon}_2 = (-0.0670, 0.4927, -0.8678)$~\cite{Gelmini:2000dm}.

  The annual variation of the scattering rate due to $\vec{v}_e(t)$ is very
nearly sinusoidal and we estimate the amplitude of the modulated
rate to be
\beq
S \equiv \left.\frac{dR}{dE_R}\right|_{mod} \simeq
\frac{1}{2}\left[
\frac{dR}{dE_R}(June\;2)-\frac{dR}{dE_R}(Dec\;2)\right].
\eeq
When computing the unmodulated rate for a given experiment, we integrate
the total rate over the time periods that the experiment
recorded data. To convert rates to detector signals, we rescale
by the efficiency of the detector and account for \emph{quenching},
but we do not include any detector resolution effects. In addition,
we have included a correction for channeling effects at DAMA, but
find them to be unimportant for the range of parameters we
consider.\footnote{The fraction of channeled events falls quickly
with recoil energy, and can be approximated for iodine as $f_I
\simeq 10^{-\sqrt{E_R/(11.5\kev)}}$~\cite{Bernabei:2007hw,Savage:2008er}.
Since IDM suppresses scattering processes with low $E_R$, one expects
only a very small number of channeled events.}

  To determine the extent to which IDM can account for the DAMA
signal, we compute the modulated recoil spectrum of IDM candidates
and compare this to the energy spectrum in the twelve lowest bins in
the $2\!-\!8\,\kevee $ range reported by DAMA in
Ref.~\cite{Bernabei:2008yi}. For a given dark matter particle mass,
we use a $\chi^2$ \emph{goodness-of-fit} metric to determine the $90\%$ and
$99\%$ confidence level allowed regions. We define this metric as
\beq
\chi^2 \equiv \sum_{i = 1}^{12} \frac{\left( S^i - S^i_{data}
\right)^2 } {\left(\sigma^i_{data}\right)^2},
\eeq
where $S^i$
denotes the average of the left, center, and right values of $S$ in
the $i^{th} ~0.5\,\kev$ width bin, and $\sigma^i_{data}$ the reported
uncertainty in the measurement.  For each value of the dark matter
mass, we scan over 2 parameters (the overall nucleon cross section
$\sigma_n^0$, and the mass splitting $\delta$), so we require that
$\chi^2 < 16.0\,(23.2)$ at the $90\,(99)\%$ level for $12 - 2 = 10$
degrees of freedom. We only consider signals from the scattering off
of iodine, as it is expected to completely dominate for the
parameters of interest to us.  Following Ref.~\cite{Chang:2008gd},
we take the quenching factor for iodine to be $q_I = 0.085$.

  As well as fitting to the DAMA results, we also compute the
signals that each IDM candidate would produce at the CDMS
experiments~\cite{Akerib:2004fq,Akerib:2005kh,Ahmed:2008eu},
CRESST-II~\cite{Angloher:2008jj}, and
ZEPLIN-III~\cite{Lebedenko:2008gb}.
This imposes further constraints on the properties of a potential
IDM candidate.\footnote{As in the analysis of
Ref.~\cite{Chang:2008gd}, we find that the constraints from other
experiments such as XENON10~\cite{Angle:2007uj},
KIMS~\cite{Lee.:2007qn}, and ZEPLIN-II~\cite{Alner:2007ja} are
currently not as important so we have not included them in our
plots.} In order to be conservative in excluding IDM parameter
space, we will assume that the small number of events seen by
these experiments are signal events, and use Poisson statistics to
find the region of parameter space excluded at the $99\%$ confidence
level based on the number of observed events.

  CDMS has published data from three runs at the Soudan
Underground Laboratory with
approximate exposures of 19.4 kg-day~\cite{Akerib:2004fq},
34 kg-day~\cite{Akerib:2005kh}, and 121.3 kg-day~\cite{Ahmed:2008eu}.
In total, these experiments reported two events between $10 \kev$
and $100 \kev$, which we assume to be signal.  We thus require that
the expected total number of events, integrated over the time
the experiments ran, obeys $N_{tot} < 8.4$ at the $99\%$ confidence level.
We only consider scattering off germanium as it is expected to dominate for
the heavier dark matter we consider.

  CRESST-II has published data from a run in 2004 using prototype
detector modules~\cite{Angloher:2004tr} and more recently has
published results from a commissioning run carried out in
2007~\cite{Angloher:2008jj}.  Since there were significant changes
to the detector modules between these runs, including the
addition of neutron shielding, we opt not to combine these data
sets and consider constraints only from the later commissioning run,
which had an exposure of 47.9 kg-day and an acceptance of 0.9 for
tungsten recoils.\footnote{We note that the combined data sets would allow parameter points
that are ruled out according to the later commissioning run
considered in isolation due to the large number of observed events in the earlier run (5)
relative to its exposure (20.5 kg-day).}
Taking the seven observed events between $12 \kev$ and $100 \kev$ to be signal,
we require that the total number of predicted tungsten recoil events, integrated
over the duration of the experiment, obeys $N_{tot} < 16.0$ at the $99\%$
confidence level.

  The ZEPLIN-III experiment has recently
released data from a run in 2008 with an effective exposure of
126.7 kg-day~\cite{Lebedenko:2008gb}. The experiment observed seven
events in its liquid xenon detector between 2~keVee and 16~keVee,
so we require that
the total number of events obeys $N_{tot} < 16.0$ at the $99\%$
confidence level. To convert measured energy $E_d$ to recoil
energy $E_R$, we use the energy-dependent quenching factor given
in Fig.\,15 of Ref.~\cite{Lebedenko:2008gb}, which saturates at
$q_{Xe} \simeq 0.48$ around $E_d = 10 \kevee$.
Below this scale, we use the parametrization $q_{Xe} \approx (0.142 ~E_d +
0.005)~\text{Exp}[-0.305 ~E_d^{~0.564}]$ given in Eq.\,(4.3) of
Ref.~\cite{MarchRussell:2008dy}, which we find gives a good fit
to the curve.

\begin{figure}[htp]
\begin{center}
        \includegraphics[totalheight=0.9\textheight]{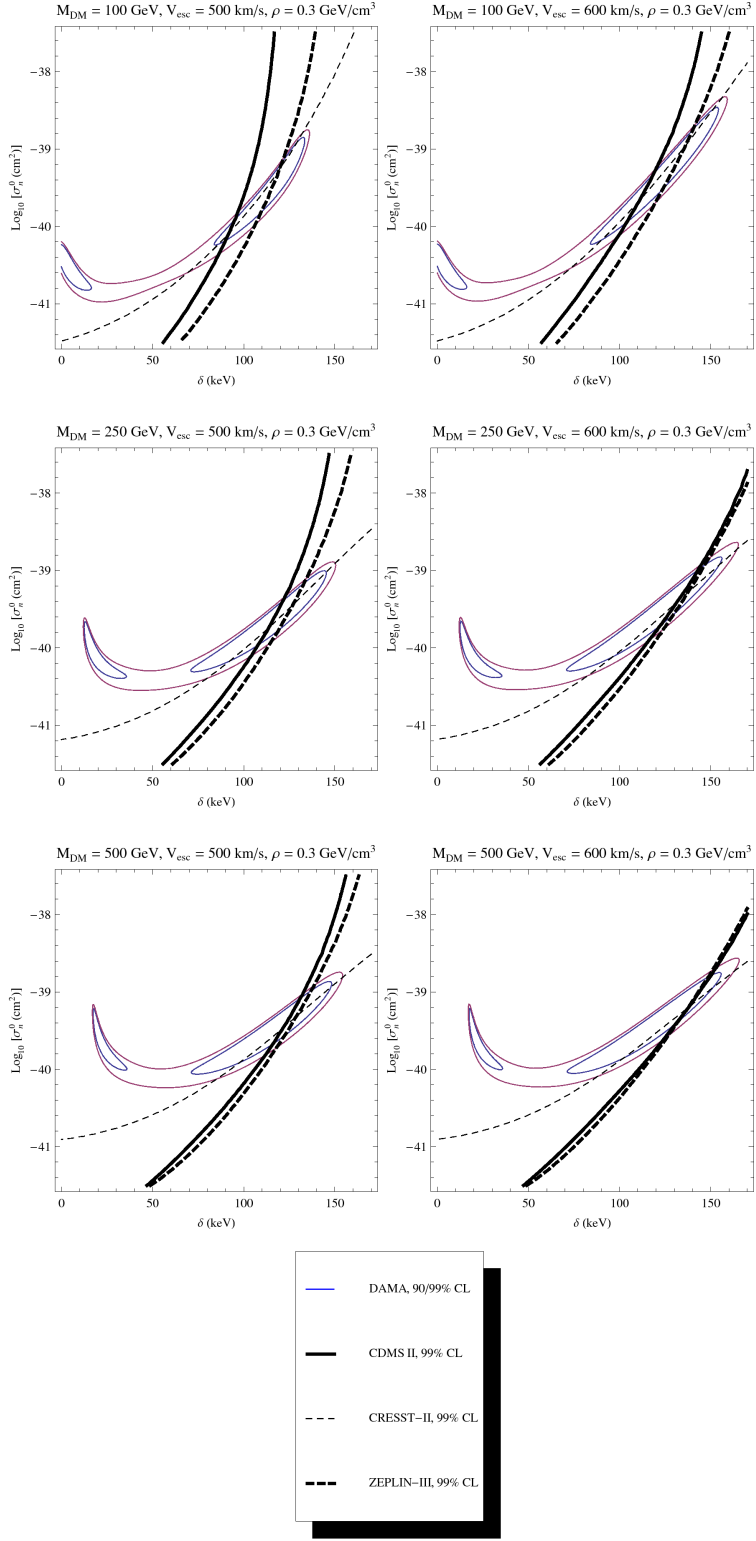}
\end{center}
\caption{Allowed regions for fits of inelastic dark matter to
the DAMA data, as well as constraints from CDMS~II, CRESST-II,
and ZEPLIN-III.  We fix the local DM density at
$\rho = 0.3 \gev/cm^3$, and vary the DM mass and
escape velocity. These plots assume the relation
$f_p = f_n$.}
\label{damafit1}
\end{figure}

\begin{figure}[htp]
\begin{center}
        \includegraphics[totalheight=0.9\textheight]{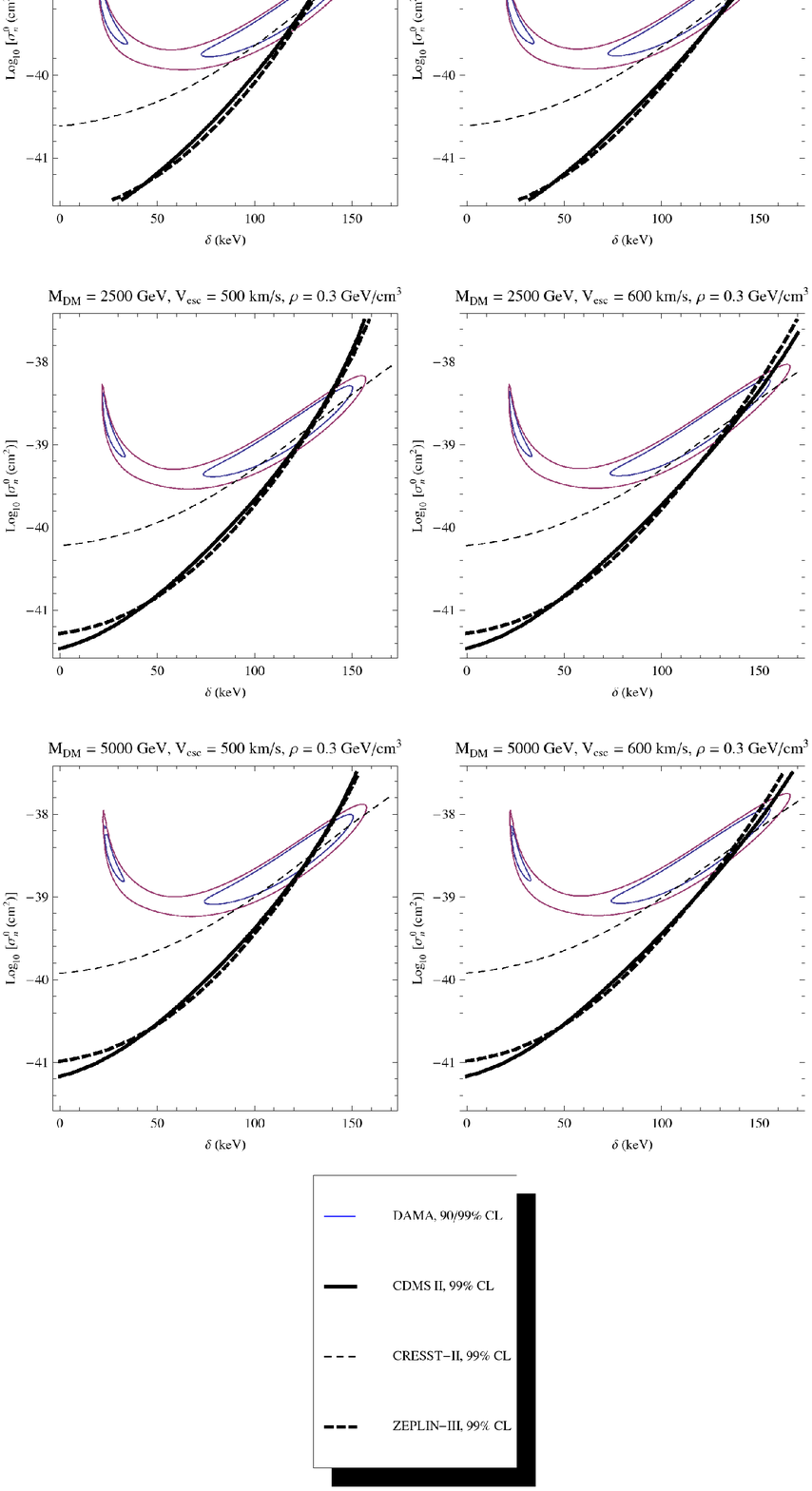}
\end{center}
\caption{Allowed regions for fits of inelastic dark matter
to the DAMA data, as well as constraints from CDMS~II, CRESST-II,
and ZEPLIN-III.  We fix the local DM density at $\rho = 0.3 \gev/cm^3$,
and vary the DM mass and escape velocity. These plots assume
the relation $f_p = f_n$.}
\label{damafit2}
\end{figure}

  In Figs.~\ref{damafit1} and \ref{damafit2} we show the region
allowed by DAMA and constraints from CDMS~II, CRESST-II, and
ZEPLIN-III at the fiducial point $f_n = f_p$ for various values of
the dark matter mass between $100\gev$ and $5\tev$. (Note that
ultimately the relation between $f_n$ and $f_p$ is
model-dependent, but we make this choice to allow a
straightforward comparison to previous work~\cite{Gelmini:2004gm,Petriello:2008jj,TuckerSmith:2001hy,TuckerSmith:2002af,TuckerSmith:2004jv,Chang:2008gd}.)
In addition, we consider escape velocities of $500 ~km/s$ and $600~km/s$.
We observe that while the relative constraints from
CDMS II, CRESST-II, and ZEPLIN-III  become stronger for
heavier dark matter candidates, very heavy dark matter is not
ruled out.  While both the required cross section for DAMA and the
constraint curves move upwards for heavier dark matter, they
 do so at roughly the same rate. One can understand this
asymptotic behavior as follows. For large values of the dark
matter mass relative to the mass of the nucleus, the reduced mass
in Eq.~\eqref{vmin} becomes $\mu_N \simeq m_N$ and is independent
of the dark matter mass.  Therefore the only mass dependence is in the
prefactor of Eq.~\eqref{rateeq} which is the same for all experiments,
so the constraint curves from different experiments do not move relative
to each other.

  It is quite interesting that heavier dark matter may be
allowed -- this opens up the possibility of simple dark matter
models for which heavier masses are preferred for getting the
right thermal relic abundance, or for explaining indirect signals
such as the $e^+ + e^-$ excess at $\sim 300 - 800 \gev$ seen by
ATIC and PPB-BETS. Models addressing the hierarchy problem also
often prefer heavier dark matter candidates ~\cite{Ponton:2008zv,ArkaniHamed:2006mb}.

  We stress, however, that the results of the present
section depend on a number of astrophysical and nuclear
physics quantities that are not fully understood and have large
uncertainties. The size of the modulated signal at DAMA is
particularly sensitive to the local dark matter velocity
distribution. For example, as one can see from Figs.~\ref{damafit1}
and \ref{damafit2}, increasing the halo escape velocity from
$500\,km/s$ to $600\,km/s$ significantly tightens the constraints
from CDMS. This is because inelastic DM scattering requires one
to sample from the tail of the velocity distribution, especially for
lighter target elements.  Deviations from the assumed Maxwellian
distribution at high velocities, as well as additional DM
substructures such as streams or sub-halos, can also
significantly affect the allowed
region~\cite{Green:2003yh,Hansen:2005yj,Vogelsberger:2008qb,
MarchRussell:2008dy}.
For this reason, we have been extremely conservative in identifying 
the excluded region by
using a \emph{goodness-of-fit} estimator and only showing $99\%$ confidence
level exclusion contours for CDMS~II, CRESST-II, and ZEPLIN-III.
Given the uncertainties, it would be premature to rule out a broader range
of IDM parameter space.

  A second important uncertainty is the value of the local dark matter
density $\rho_{DM}$, which is known only up to a factor of $\sim 2$.
Varying $\rho_{DM}$ will affect the overall normalization of the
cross section $\sigma_n^0$ in the plots of Figs.~\ref{damafit1} and
\ref{damafit2} by a factor inversely proportional to it.  While this
is not terribly important for demonstrating the existence of an
allowed region, since it affects all signals equally, the
uncertainty in $\rho_{DM}$ can be very important when comparing the
allowed region to a model that  predicts a specific value for
$\sigma_n^0$.

  Finally, we wish to emphasize that the analysis performed in this
section (and in Refs.~\cite{Chang:2008gd,Savage:2008er,MarchRussell:2008dy})
is not model-independent. Simple models of IDM, for example dark matter
charged under $SU(2)_L$, will not respect the relation $f_p = f_n$,
and the curves in Figs.~\ref{damafit1} and \ref{damafit2}
will move around by different amounts that depend on the
atomic numbers relevant to the experiment. In
Section~\ref{general}, we will redo the analysis for dark matter
charged under $SU(2)_L$, which scatters through $Z^{0}$-exchange.
There we will see that when one considers the heavier dark matter
masses that yield the correct thermal relic abundance
\cite{Cirelli:2005uq}, the scattering cross section from
$Z^{0}$-exchange is tantalizingly close to the DAMA preferred region
(which remains qualitatively similar to the $f_p=f_n$ case considered here).

\subsection{Other DM Explanations for DAMA}

  In addition to IDM, a number of other non-standard DM candidates
have been suggested as possible sources for the DAMA signal.
These include lighter spin-independent elastic DM~\cite{Gelmini:2004gm},
lighter spin-dependent elastic DM~\cite{Savage:2004fn},
and elastic scattering off atomic electrons~\cite{Bernabei:2007gr}.  These
alternatives to IDM turn out to be extremely strongly constrained when
spectral data from DAMA is included in the
analysis~\cite{Chang:2008xa,Fairbairn:2008gz,Savage:2008er}.

  Light $2\!-\!10\,\gev$ elastic, spin-independent DM was proposed
as an explanation for the DAMA signal in Ref.~\cite{Gelmini:2004gm}.
The DAMA signal in this scenario comes primarily from the light DM
scattering off sodium rather than iodine. Such a light DM state would produce
recoils near the lower end of the sensitivity range of most other
direct detection experiments which consist of heavier target nuclei,
strongly suppressing their signals relative to DAMA. The effect of
light DM on DAMA can be further enhanced relative to other direct
detection probes by \emph{channeling}, which effectively reduces the
amount of quenching in the DAMA target, leading to more events in
the range of sensitivity~\cite{Petriello:2008jj}.

However,
subsequent analyses taking into account the modulated and
unmodulated single-hit DAMA energy spectra indicate that this light
DM scenario is strongly disfavored. It is found that light DM either
does not provide a good fit to the modulated spectrum or predicts
an unmodulated single-hit rate in the lowest energy bins that is
much larger than the total rate observed by
DAMA~\cite{Chang:2008xa}. Even so, light spin-dependent DM may still
be viable if channeling is included~\cite{Savage:2008er}.

  A second alternative explanation for the DAMA result consists
of moderately heavy ($m_{DM} \gtrsim 10\,\gev$) elastic DM that
scatters primarily off atomic electrons rather than
nuclei~\cite{Bernabei:2007gr}. The vast majority of these
scatterings would produce an electromagnetic signal in the DAMA
detector in the eV energy range, well below the keV energies to
which this detector is sensitive. However, the scattering of a
halo DM particle off an atomic electron with an unusually high
momentum in the tail of its distribution, on the order of an MeV,
can generate a detected electromagnetic signal at DAMA of $E_d
\sim \mbox{few}\,\kev$~\cite{Bernabei:2007gr}. No such signal
would have been recorded in other direct detection experiments
such as CDMS and XENON since these experiments are careful to
filter out electromagnetic events that they expect to arise from
backgrounds. Electron scattering DM is also attractive in light of
the PAMELA results~\cite{Bai:2008jt,Fox:2008kb}, which can be
interpreted as coming from DM annihilating preferentially into
leptons~\cite{Cirelli:2008pk,Cholis:2008wq}.

  Following the analysis of Ref.~\cite{Chang:2008xa} for light DM
scattering off nuclei, we investigate whether electron
interacting DM is consistent with the modulated and unmodulated
single-hit DAMA energy spectra. In our analysis we compute the
modulated and unmodulated rates for DM scattering off electrons as
in Ref.~\cite{Bernabei:2007gr}. We compare our binned results to
the lowest twelve $2\!-\!8\,\kevee$ DAMA modulated bins and the
lowest six $0.875\!-\!2.125\,\kevee$ DAMA unmodulated single-hit
bins using a modified $\chi^2$ measure.  We use a standard
$\chi^2$ \emph{goodness-of-fit} measure for the modulated bins,
while for the unmodulated bins we add to the $\chi^2$ only if the
predicted signal is larger than the observed value to allow for an
unmodulated background. This procedure is very conservative in
that it will only underestimate the excluded regions.

  We find that under the assumptions about the DM
made in Ref.~\cite{Bernabei:2007gr}, namely that the DM
is fermionic and interacts with quarks by the exchange of
a scalar or a gauge boson with $(V \pm A)$
couplings,\footnote{We also neglect parts of the
electron-DM cross section suppressed by the DM velocity.}
electron-interacting DM as an explanation for DAMA is excluded
well beyond the $99\%$ confidence level.
This occurs for precisely the same reasons that light elastic DM
is strongly disfavored: either the modulated signal is too low,
or the unmodulated single-hit signal (\emph{i.e.}, the signal
excluding multiple scintillation events) exceeds the total rate
observed by DAMA.  In the present case, the signal rate falls
quickly with increasing detected energy $E_d$ because the
momentum distribution of atomic electrons decreases rapidly
in the relevant range, approximately as $p^{-8}$~\cite{Bernabei:2007gr}.

\begin{figure}[ttt]
\vspace{1cm}
\begin{center}
        \includegraphics[width = 0.7\textwidth]{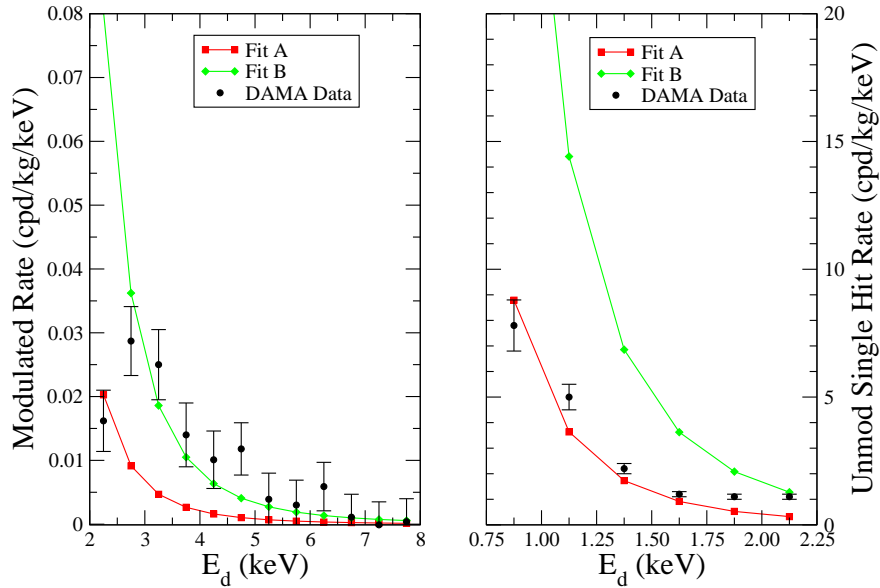}
\end{center}
\caption{DAMA modulated and unmodulated single hit spectral data
and fits of electromagnetic elastic DM scattering.  The DM particle
is assumed to be a fermion of mass $M_{DM} = 200\,\gev$
with either scalar or chiral vector ($V\pm A$) couplings to electrons.
The (red) curve for Fit~A is the one that yields the lowest
effective $\chi^2$ value using both the modulated and single-hit
unmodulated spectral data sets.  The (green) curve for
Fit~B corresponds to that with the lowest effective $\chi^2$
for the modulated dataset alone, excluding the lowest energy bin.}
\label{dmemfits}
\end{figure}

  This tension is illustrated in Fig.~\ref{dmemfits} where
we show the best fit (lowest effective
$\chi^2$) to the full DAMA spectral dataset (Fit A),
as well as the best fit (lowest effective $\chi^2$) to the modulated
dataset alone leaving out the lowest $2\,\kevee$ energy bin~(Fit B).
In making these fits, we assume either chiral vector
($V\pm A$) or scalar four-fermion interactions between the electron and
a fermionic DM particle of mass equal to $200\,\gev$.
However, the shape of the predicted spectrum,
which is the source of the tension with the DAMA data,
is effectively independent of the DM mass provided it
is heavier than about $10\,\gev$~\cite{Bernabei:2007gr}.
Therefore the curves in Fig.~\ref{dmemfits} also apply to other DM masses
provided we rescale the effective DM-electron coupling strength appropriately.

  Our conclusion that electron-DM scattering gives a poor fit
to the DAMA spectral data is robust.  We find that
it continues to hold even if we do not include the lowest modulated
energy bin and the two lowest unmodulated energy bins in the fit.
Furthermore, we have also examined other Dirac structures for the
couplings between a fermionic DM particle and the target electron
(relative to the scalar and vector $(V\pm A)$ couplings
considered above and in Ref.~\cite{Bernabei:2007gr}),
and we find that these do not improve the situation.
A scalar DM particle scattering elastically off electrons
does not appear to work either.\footnote{
Inelastic dark matter scattering off electrons might work,
although this would likely require an extremely large electron
scattering cross section.}

\section{General IDM Properties and Nucleon Scattering
\label{general}}

  Having found that IDM provides an acceptable fit to the DAMA
modulated and unmodulated spectral datasets and that other
proposals are strongly constrained, we turn now to the general
properties of potential IDM candidates.  Clearly, any such candidate
must have an inelastic nucleon-scattering cross section that is
significantly enhanced relative to its elastic
cross section. To meet these criteria,  the mass, inelastic
splitting, and inelastic cross section must all fall into the
appropriate ranges. We now show these requirements can be satisfied
by IDM particles in the mass range $100\gev\!-\!5\tev$ that interact
coherently with nuclei primarily through the exchange of a massive
gauge boson.

\subsection{Inelastic Interactions from a Massive Gauge Boson}

  When the DM-nucleon scattering is mediated by a massive gauge boson,
the dominance of inelastic interactions over elastic arises
in a natural way.  We consider two types of  IDM in this context:
a Dirac fermion that is split into a pair of nearly degenerate
Majorana states by a small Majorana mass, and a complex scalar that
is split into two real scalars by a small {holomorphic}
mass.\footnote{Another interesting possibility would be spin 1
inelastic dark matter, but models of this type are  more complicated.  }

  For the fermion case, the models we consider all reduce to
\beq
\mathscr{L} \supset
\bar{\psi}i\gamma^{\mu}(\del_{\mu}+igQ\,Z_{\mu}')\psi
- M\bar{\psi}\psi
- \frac{1}{2}m_L\left(\bar{\psi^c}P_L\psi + h.c.\right)
- \frac{1}{2}m_R\left(\bar{\psi^c}P_R\psi + h.c.\right),
\label{fermidm}
\eeq
where $\psi$ is a Dirac fermion, $M\gg m_{L,R}$, and $Z_{\mu}'$ is a
massive gauge boson.  The Majorana masses $m_{L,R}$
in Eq.~\eqref{fermidm} split the Dirac state $\psi$ into
a pair of Majorana states $\Psi_{1,2}$.
In terms of these mass eigenstates, the Lagrangian becomes

\bea \mathscr{L} &\supset&
\frac{1}{2}\bar{\Psi}_1i\gamma^{\mu}\del_{\mu}\Psi_1 -
\frac{1}{2}(M-m_+)\bar{\Psi}_1{\Psi}_1
\\
&& + \frac{1}{2}\bar{\Psi}_2i\gamma^{\mu}\del_{\mu}\Psi_2 -
\frac{1}{2}(M+m_+)\bar{\Psi}_2{\Psi}_2
\nnmb\\
&& + i\,g\,Q\,Z_{\mu}'\,\bar{\Psi}_2\gamma_{\mu}\Psi_1
\nnmb\\
&& + \frac{1}{2}\,g\,Q\,Z_{\mu}'\,\frac{\phantom{.}m_-}{M}\,\left(
\bar{\Psi}_2\gamma^{\mu}\gamma^5\Psi_2
-\bar{\Psi}_1\gamma^{\mu}\gamma^5\Psi_1\right) +
\mathcal{O}\lrf{m^2}{M^2}, \nnmb \eea where $m_{\pm} = (m_L\pm
m_R)/2$.

From this we see that the dominant gauge boson
interaction is strictly off-diagonal, and that the mass splitting
between the eigenstates is \beq \delta = M_2-M_1 = 2\,m_+ =
m_L+m_R. \eeq There is also a residual diagonal coupling of the
fermions to the gauge boson, but it is suppressed by a power of
$m_-/M \ll 1$.

  The basic story for the scalar case is very similar.
Consider the interactions
\beq
\mathscr{L} \supset |(\del_{\mu}+igQ\,Z_{\mu}')\phi|^2 - M^2|\phi|^2
-\frac{1}{2}m^2(\phi^2+h.c.),
\label{scaldm}
\eeq
where, again, $Z_{\mu}'$ is a massive vector boson and we assume $M^2 \gg m^2$
with $m^2$ real and positive.  The holomorphic $m^2$ mass term splits
the real and imaginary components of the complex scalar
$\phi = (\phi_R+i\phi_I)/\sqrt{2}$.  In terms of these fields,
the Lagrangian becomes
\bea
\mathscr{L} \supset \frac{1}{2}(\del\phi_R)^2-\frac{1}{2}(M^2+m^2)\phi_R^2
+\frac{1}{2}(\del\phi_I)^2-\frac{1}{2}(M^2-m^2)\phi_I^2\\
- gQZ^{' \mu}(\phi_I\del_{\mu}\phi_R-\phi_R\del_{\mu}\phi_I)
+\frac{1}{2}g^2Q^2Z_{\mu}'Z^{' \mu}(\phi_R^2+\phi_I^2) . \nnmb \eea
From this, we see that the  single gauge boson interaction with two
scalars is strictly off-diagonal, coupling $\phi_R$ exclusively to
$\phi_I$. The splitting between these mass eigenstates is \beq
\delta = \sqrt{M^2+m^2}-\sqrt{M^2-m^2} \simeq \frac{m^2}{M}. \eeq

  Let us also emphasize that in both the fermion and scalar
IDM cases, it is technically natural to have the ``Dirac'' mass
$M$ much larger than the ``Majorana'' mass $m$. In the limit $m\to
0$, both theories have a global $U(1)_{DM}$ symmetry analogous to
baryon number in the SM, implying that all quantum corrections to
$m$ (or $m^2$) are proportional to itself.  Indeed, within the
MSSM the $B$ and $L$ global symmetries keep the real and imaginary
components of the squarks and sleptons degenerate, while the
VEVs of the Higgs complex scalars split their components.
In Section~\ref{models}, we will construct models that generate such
inelastic splittings in simple and natural ways.

\subsection{Nucleon Scattering Rates}

  To compute the effective nucleon scattering rates $\sigma^0_{p,n}$
relevant for Eq.~\eqref{dsigder} mediated by a massive gauge boson,
we concentrate exclusively on vector-vector (VV) interactions.
Such VV interactions give rise to coherent \emph{spin-independent}
DM scattering off target nuclei~\cite{Jungman:1995df}.
Axial-axial (AA) interactions, on the other hand, produce
an incoherent \emph{spin-dependent} coupling to target nuclei.
Mixed VA and AV interactions can also be neglected because they
produce effective scattering cross sections suppressed by at least
two powers of the DM velocity, which is on the order of
$v\sim 10^{-3}$ in our galactic halo.

  For fermionic IDM arising from couplings of the form given in
Eq.~\eqref{fermidm}, the effective nucleon cross section
$\sigma_n^0$ needed to compute the interaction rate
Eq.~\eqref{dsigder} is identical to the cross section for a Dirac
fermion to scatter off the nucleon. Up to small corrections, the
effect of the inelasticity is completely accounted for by setting
the lower velocity cutoff $v_{min}$ in Eq.~\eqref{rateeq} to the
expression given in Eq.~\eqref{vmin}. Starting with the Lagrangian
for vector couplings of the SM quarks $q$ and a Dirac fermion
$\psi$ to a massive $Z'$ gauge boson (note we are using $Z'$ here
for generality -- the next subsection will restrict attention to
the Standard Model $Z^0$),
\beq
\mathscr{L} \supset
-g\,g_V^q\,Z_{\mu}'\,\bar{q}\gamma^{\mu}q -
g\,g_V^{\psi}\,Z_{\mu}'\,\bar{\psi}\gamma^{\mu}\psi,
\eeq
the relevant effective nucleon-scattering cross section
is~\cite{Jungman:1995df}
\beq
\sigma_{p,n}^0 =
\frac{1}{\pi}\mu_{p,n}^2\lrf{g}{M_{Z'}}^4 (g_V^{\psi}g_V^{p,n})^2,
\label{sigmaferm}
\eeq
where $\mu_{p,n}$ is the reduced mass of
the $\psi$-nucleon system, and
\beq g_V^p = 2\,g_V^u +
g_V^d,~~~~g_V^n = g_V^u + 2\,g_V^d.
\eeq
The couplings $f_p$ and $f_n$ appearing in Eq.~\eqref{dsigder}
coincide with $g_V^p$ and $g_V^n$ in the present case.

  For complex scalar IDM, the effective nucleon scattering cross section
required to evaluate the event rate in Eq.~\eqref{rateeq} is
identical to the cross section for a single Dirac fermion of the same mass
as the scalar to scatter off a nucleon. Again the effects of the inelasticity are
accounted for by modifying the lower velocity cutoff $v_{min}$ in
Eq.~\eqref{rateeq}. With the coupling of a complex scalar to a
massive $Z'$ given by
\beq
\mathscr{L} \supset - i
g\,g_V^{\phi}\,{Z'}^{\mu}\,(\phi^* \del_{\mu} \phi - \phi
\del_{\mu} \phi^*),
\eeq
 the effective nucleon scattering cross section is
\beq
\sigma_{p,n}^0 = \frac{1}{\pi}\,\mu_{p,n}^2\,\lrf{g}{M_{Z'}}^4\,
(g_V^{\phi}g_V^{p,n})^2.
\label{sigmascal}
\eeq
As for the fermionic case, we can identify $g_V^{p,n}$ with $f_p$ and $f_n$
appearing in Eq.~\eqref{dsigder}.

\subsection{$SU(2)_L$ Mediation}

  Perhaps the simplest possibility to mediate IDM scattering off
nucleons is the $Z^0$ gauge boson of the SM.  Note that the photon
is not an option because we assume that the DM candidate is neutral and
$W^{\pm}$ is not an option because the radiatively induced mass
splitting between charged and neutral components of a multiplet
scales as $\alpha_W M_W \sim 100 \mev$, which is much too large
for IDM. A neutral particle that couples to the $Z^0$ necessarily
carries hypercharge,  so the simplest possibility is that dark
matter is a doublet of $SU(2)_L$ with hypercharge $Y = 1/2$.

  For a Dirac fermion or a complex scalar $SU(2)_L$ doublet with
hypercharge 1/2, the effective couplings for the neutral
components are $g_V^{DM} = {1}/{2}$,  $g_V^p =
{1}/{4} - \sin^2\theta_W$, and $g_V^n = -1/4$. Since
$\sin^2\theta_W \simeq 0.24$, the neutron coupling is much larger
than the proton coupling. The corresponding cross section is
\beq
\sigma_n^0 = \frac{G_F^2}{2\pi}\mu_n^2 \simeq 7.44\times
10^{-39}\,cm^2.
\eeq
To obtain this number, we assumed that the DM mass is much
larger than that of the neutron so that $\mu_n \simeq m_n = 0.9396\,\gev$.

\begin{figure}[htb]
\vspace{1cm}
\begin{center}
        \includegraphics[width = 0.7\textwidth]{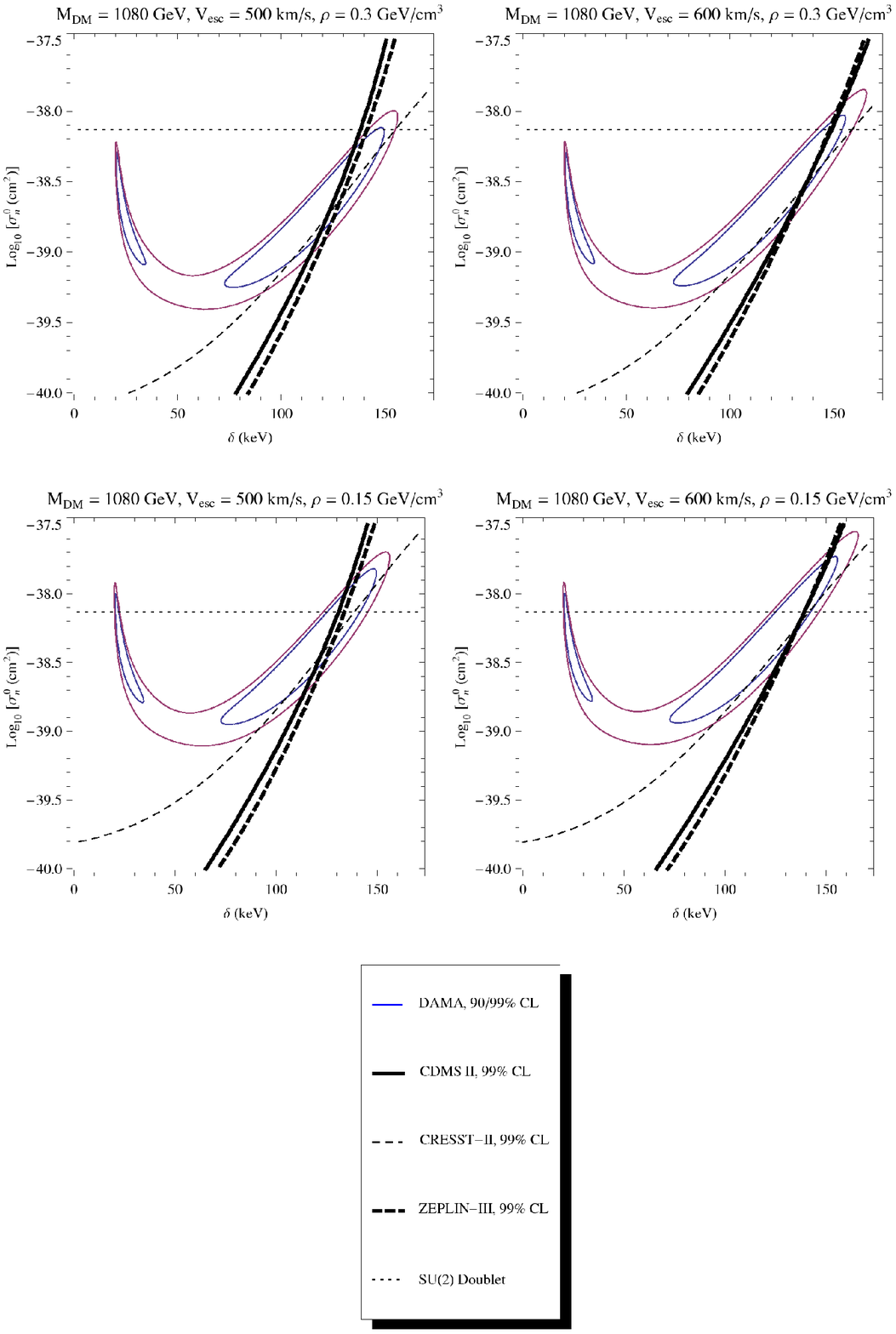}
\end{center}
\caption{DAMA allowed region and constraints for scattering
through $Z^0$-exchange for a fermion $SU(2)_L$ doublet with a mass
of $1080\,\gev$, a value that yields the correct thermal relic density.
We consider values of the DM escape velocity of $v_{esc}
= 500\,km/s$ and $600\,km/s$ and local DM densities
$0.15\,\gev/cm^3$ and $0.3\,\gev/cm^3$.}
\label{zplotsfermion}
\end{figure}

\begin{figure}[htb]
\vspace{1cm}
\begin{center}
        \includegraphics[width = 0.7\textwidth]{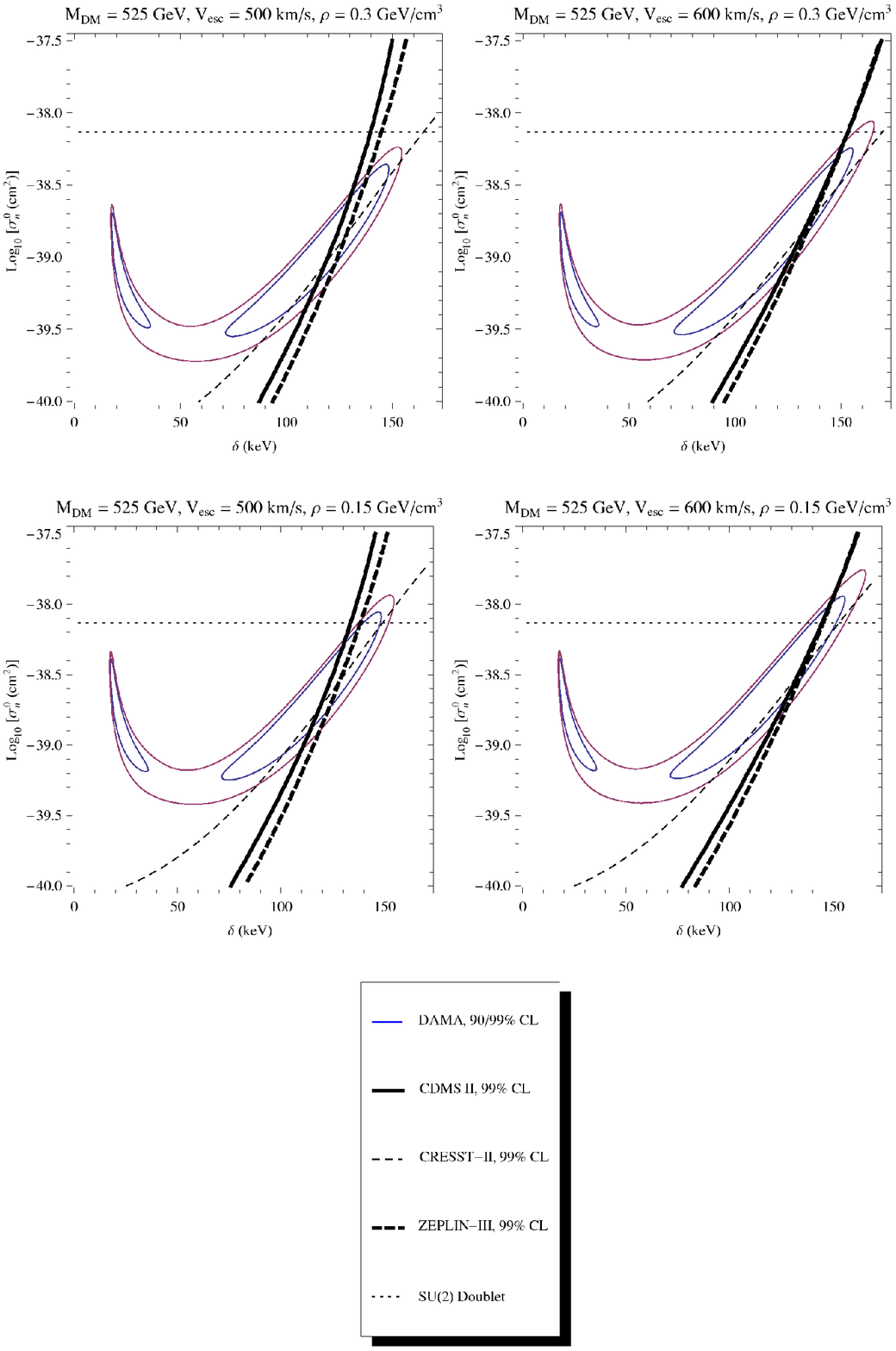}
\end{center}
\caption{DAMA allowed region and constraints for scattering
through $Z^0$-exchange for a scalar $SU(2)_L$ doublet with a mass
of $525\,\gev$, a value that yields the correct thermal relic
density. We consider values of the DM escape velocity of $v_{esc}
= 500\,km/s$ and $600\,km/s$ and local DM densities
$0.15\,\gev/cm^3$ and $0.3\,\gev/cm^3$.} \label{zplotsscalar}
\end{figure}

  In Figs.~\ref{zplotsfermion} and \ref{zplotsscalar} we show fits
to the DAMA modulated data along with constraints from CDMS II,
CRESST-II, and ZEPLIN-III, for nucleon scattering mediated by
$Z^0$-exchange and DM masses of $1080\,\gev$ and $525\,\gev$,
respectively.  These particular masses were chosen because they
are the values that lead to the correct thermal relic density for
a Dirac fermion ($1080\,\gev$) and a complex scalar doublet
($525\,\gev$)~\cite{Cirelli:2005uq}.\footnote{ Using micrOMEGAs
v2.2~\cite{Belanger:2006is}, we find slightly smaller central
values for the preferred masses of $1080\,\gev$ and $525\,\gev$
as compared to Ref.~\cite{Cirelli:2005uq}, who obtain $1100\,\gev$
and $540\,\gev$.  However, these differences are within the margin of error
and have a very small effect on the allowed region where the relic density
scales roughly as $\Omega_{DM}h^2 \propto m_{DM}^2$.}

It is intriguing that the inelastic cross section
mediated by $Z^0$ exchange is very similar to the values preferred
by our fit to DAMA. This coincidence of scales was observed in
Ref.~\cite{TuckerSmith:2004jv} in fitting the case of mixed
sneutrino DM to the DAMA/NaI dataset. Here we observe  that this persists
 even with more detailed fits to
the energy spectrum and for more general doublet candidates.

  The $Z^0$-mediated nucleon cross section is in fact a little bit
too big assuming a local DM density of $\rho_{DM} =
0.3\,\gev/cm^3$. However, there is a significant uncertainty in
the local DM density and lowering its value to $\rho_{DM} =
0.15\,\gev/cm^3$, which is also within the allowed range, leads to
good agreement at the $90\%$ confidence level for fermion doublet
DM, and to marginal agreement for scalar doublet DM.  Another
possibility is that the dark matter has multiple
components~\cite{Hur:2007ur,Feng:2008ya,
Fairbairn:2008fb,Zurek:2008qg}, with the local density of the
doublet component giving rise to the DAMA signal well below
$0.3\,\gev/cm^3$.  The fermion or scalar doublet DM could also be
heavier than $1080\,\gev$ or $525\,\gev$ and its density diluted
by a late-time production of entropy, allowing for a larger
cross section.

  Let us also emphasize that the mass splitting terms in
Eqs.~(\ref{fermidm}) and \eqref{scaldm}
necessarily break $U(1)_Y$.  However, such a mass
term can be generated after electroweak symmetry breaking through operators
involving the Higgs field~\cite{Hall:1997ah}. The exact
operator one needs then depends on the quantum numbers of the dark
matter particle. For $SU(2)_L$-doublet dark matter with
hypercharge $Y = 1/2$, we can introduce the gauge invariant operator
\beq
\label{scalarop}
\mathscr{L} \supset -\frac{\lambda}{2} (
\phi \phi h h + h.c. )
\eeq
for a scalar, or
\beq
\label{fermop}
\mathscr{L} \supset -\frac{1}{2 \Lambda} (\bar{\psi}^c \psi h h
+ h.c.)
\eeq
for a fermion. In order to obtain a splitting $\delta
\sim 100 \kev$, we need $\lambda \sim (\delta / v)(m_{\phi} / v)
\sim 10^{-6}$, or $\Lambda \sim (v^2 / \delta) \sim 10^8 \gev$.
While without further model structure (such as we will soon consider) these are somewhat awkward
numbers, we emphasize that the values are technically natural.

  Of course one can consider  representations aside from an
$SU(2)_L$ doublet.  The next simplest possibility is to introduce
a complex triplet of $SU(2)_L$ with hypercharge $Y = 1$. In order
to split the states that couple to the $Z^0$, the triplet would
require a higher-dimension operator $(T^{i j} h_i h_j)^2$ in order
to be gauge invariant.  For scalar DM, this operator should be
suppressed by a scale $\Lambda^2 \sim (v^4 / m_T \delta ) \sim
(10^{5} \gev)^2$, and for fermion DM it should be suppressed by a
scale $\Lambda^3 \sim (v^4 / \delta) \sim (3\times 10^{6}
\gev)^3$.

  One can obviously keep considering larger representations of $SU(2)_L$,
which will cause the scale suppressing the smallest gauge
invariant splitting operator to decrease even further.  In
extensions to the Standard Model that solve the hierarchy problem,
one might naturally expect to have operators suppressed by the TeV
scale. Large enough representations of $SU(2)_L$ (\emph{e.g.} a
\textbf{5} or a \textbf{7}) may then naturally have the correct
splitting.  For a representation of dimension $N$, the DM
mass needed to reproduce the right relic density, and hence the
cross section needed for DAMA, increases roughly as $N^{3/2}$.
On the other hand, the effective scattering cross section $\sigma^0_n$
depends on the choice of hypercharge, and can scale
between $1$ and $N^2$.
This means that thermal dark matter composed of a higher-dimensional
representation can also agree with direct detection constraints,
though the agreement can be better or worse according to the
direct detection cross sections.

  While it would be interesting to study the fits of larger
representations of $SU(2)_L$ in more detail, we find it intriguing
that the simplest possibility of an $SU(2)_L$ doublet works
reasonably well for explaining the DAMA data.  If one is to take
this model of dark matter seriously, then, the main question is
whether the numbers and scales cited above in
Eqs.~\eqref{scalarop} and \eqref{fermop}, though technically
natural, have a reasonable origin in models.   That is, suppose
DAMA has indeed discovered dark matter. What would be a reasonable
interpretation of this result? In Section~\ref{models}, we will
consider several possibilities for explaining the physical origin
of these operators and their coefficients.

\clearpage

\subsection{$U(1)_x$ Mediation}

  Inelastic DM scattering can also be mediated by a  massive
gauge boson not existing in the Standard Model.  Two distinct possibilities that can yield the right
cross sections are a heavy (TeV-scale) gauge boson with
order unity couplings to Standard Model fields, and a light ($M_{Z'}
\ll M_{Z^0}$) hidden gauge boson with highly suppressed couplings to
the SM. If the hidden gauge symmetry is an Abelian $U(1)_x$, small
couplings arise in a natural way from kinetic mixing with hypercharge.
We consider here both the heavy and light exotic gauge boson cases,
focusing on an Abelian $U(1)_x$ gauge symmetry for
simplicity.

\subsubsection{A Heavy Visible $U(1)_x$}

 A heavy $U(1)_x$ $Z'$ gauge boson with order unity
couplings to the SM can conflict with phenomenological bounds
on the mass of this new state. Collider and other bounds
place lower limits on the $Z'$ mass, whereas a very heavy gauge
boson  generate a nuclear scattering cross section that is too
small to account for DAMA, yielding some tension in this scenario.

   Precision measurements at LEP imply that the bounds on lepton
couplings for a given $Z'$ mass are generally stronger than those
for quarks. For example, a $(B\!-\!L)$ gauge boson would not
satisfy phenomenological bounds and allow for a DAMA signal
without unreasonably large couplings to the DM particle or an
extremely small gauge coupling.
Satisfying phenomenological constraints, even for a gauge coupling
to the SM as small as $g_x \sim 0.4$, requires a gauge boson mass
$M_{Z'} \gtrsim 2.5\tev$. To achieve the DAMA signal would then
require the effective coupling to dark matter to be greater than
$g_x\,x_{DM} \gtrsim 10$, where $x_{DM}$ is the dark matter
charge.  Lighter $(B\!-\!L)$ gauge bosons with perturbative couplings
to the DM ($g_x\,x_{DM} \lesssim 1$) are possible at the expense of
making the gauge coupling $g_x \lesssim 0.05$ while keeping
$x_{DM} \gtrsim 20$.  Such a hierarchy of charges seems contrived.

\begin{figure}[ttt]
\vspace{1cm}
\begin{center}
        \includegraphics[width = 0.5\textwidth]{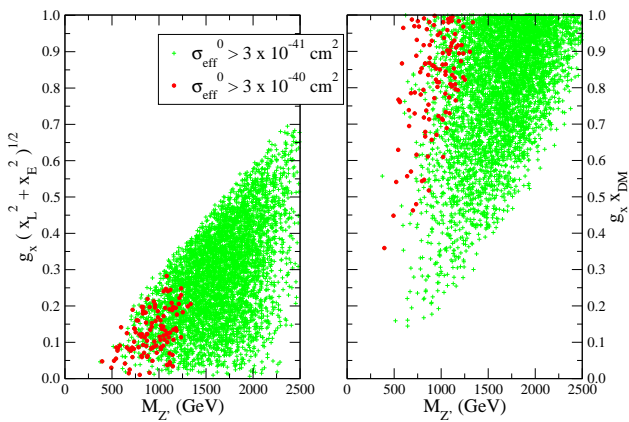}
\end{center}
\caption{Values of the lepton and dark matter charges under
$U(1)_x$ consistent with the phenomenological constraints listed
in the text that also generate an adequately large value of the
effective nucleon scattering cross section $\sigma_{eff}^0$, as a
function of the $Z'$ mass $M_{Z'}$. The charges of the quark and
Higgs fields are also scanned over.}
\label{xcharges}
\end{figure}

  Models with smaller coupings to leptons relative to quarks
and the DM are more reasonable. In Fig.~\ref{xcharges} we show the values of
the $U(1)_x$ charges of the leptons ($x_E$ and $x_L$ for $e_R^c$ and $L$)
and the DM particle ($x_{DM}$) for allowed points from a scan over
heavy $U(1)_x$ models.  These points satisfy both the phenomenological
constraints on a heavy Abelian gauge boson and generate
a reasonably large nuclear scattering cross section.
We assume flavor-independent gauge charges (consistent with Yukawa
couplings) for simplicity and to minimize flavor-violation,
and take a two-Higgs doublet model for generality.
Anomaly cancellation can be satisfied by adding exotic
fermions.  We have  focused on charges such that $g_x\,x_i \leq 1$
to ensure weak coupling.

  The phenomenological bounds we apply in generating Fig.~\ref{xcharges}
are the direct search bounds from the Tevatron in the
di-lepton~\cite{StelzerChilton:2008rx} and
di-top~\cite{Schwanenberger:2006tv} channels and the
limits on contact interactions from LEP~II~\cite{:2003ih}. In
applying the Tevatron bounds, we assume the $Z'$ decays entirely
into SM final states.  We do not explicitly compute the effects of
the new $U(1)_x$ on  precision electroweak observables, but
demand that the $Z^0\!-\!Z'$ mixing angle be less than
$\theta_{mix} < 3\times
10^{-3}$~\cite{Holdom:1990xp,Umeda:1998nq,Appelquist:2002mw,
Carena:2004xs,Kumar:2006gm,Langacker:2008yv},
which approximately captures the constraints from these
observables~\cite{Appelquist:2002mw}. In computing the mixing
angle, we assume that $\tan\beta = 10$.  For the nucleon
scattering cross section, we demand that the quantity
\beq
\sigma^0_{eff} \equiv
\sigma^0_n\,\frac{[Z\,g_V^p+(A-Z)\,g_V^n]^2}{(g_V^n)^2A^2}
\eeq
be larger than $3\times 10^{-41}\,cm^2$ or $3\times 10^{-40}\,cm^2$.
This definition and these lower limit values are motivated by our
fits to IDM in Section~\ref{fits}, where we assumed
$f_p=g_V^p=f_n=g_V^n$ in
Eq.~\eqref{dsigder}.

  We conclude from Fig.~\ref{xcharges} that a $Z'$ with a mass
in the range of $400\,\gev$ to several TeV can generate an
adequately large nucleon scattering cross section to account for
DAMA while not running afoul of the existing bounds. Such a $Z'$
must generally be somewhat leptophobic, and have significantly
large couplings to quarks and the DM state. However,
 we note these couplings are so large that the
corresponding gauge coupling generally
encounters a Landau pole well below the GUT scale, even without
including the effects of possible exotics required for anomaly
cancelation (which would only make this problem worse).
Thus, many popular $U(1)_x$ models such as
$(B\!-\!L)$ and the exotic $U(1)$'s motivated by
$E_6$~\cite{Rizzo:2006nw} do not work if
a unification relation is imposed on the exotic gauge coupling.
Furthermore, although this heavy $Z'$ scenario is a possibility,
the natural connection to the thermal relic abundance that was
present with an electroweak cross section is in general lost.

\subsubsection{A Light Hidden $U(1)_x$}

  A second possibility for an exotic Abelian gauge symmetry
mediating IDM scattering at DAMA consists of a relatively light
hidden $U(1)_x$ that couples only weakly to the SM. Very small SM
couplings arise naturally if the hidden sector couples to the SM
through kinetic mixing of the $U(1)_x$ with hypercharge.
A coupling~\cite{Holdom:1985ag}
\beq
\mathscr{L} \supset
-\frac{\epsilon}{2}\,B_{\mu\nu}X^{\mu\nu},
\eeq
where $B_{\mu\nu}$ and $X_{\mu\nu}$ are the otherwise canonically
normalized $U(1)_Y$ and $U(1)_x$ field strengths, can arise from
loops of heavy states that are charged under both gauge groups.
This leads to typical
values of $\epsilon \simeq 10^{-4}\!-\!10^{-2}$~\cite{Dienes:1996zr}.
Making a field redefinition to eliminate the kinetic mixing term,
the SM matter fields acquire effective charges under the $U(1)_x$.
If the DM states couple directly to the $U(1)_x$ but not the SM gauge
groups, the exotic gauge boson can mediate DM scattering off
nuclei.

   We are interested in the case where the exotic $Z'$ gauge boson
is light, $M_{Z'} \ll M_Z$.  In this limit the $U(1)_x$ can be
treated as mixing with $U(1)_{em}$, and the induced charges are
\beq
\mathscr{L} \supset
e\,\cos\theta_W\,\epsilon\,Q\,Z_{\mu}'\,\bar{f}\gamma^{\mu}f  +
\mathcal{O}(\epsilon\,M_{Z'}^2/M_{Z}^2),
\eeq
where $f$ represents
a SM fermion with electric charge $Q$.  Since the light $Z'$ couples
to electric charge in the visible sector, it will  mediate
scattering of the DM only with protons. Applying Eq.~\eqref{sigmaferm}
or Eq.~\eqref{sigmascal}, the corresponding cross section is
\beq
\sigma_p^0 = \lrf{g_x\,x_{DM}}{0.5}^2\lrf{\gev}{M_{Z'}}^4
\lrf{\epsilon}{10^{-3}}^2\,(2.1\times 10^{-36}\,cm^2),
\label{lightu1sig}
\eeq
where $g_x\,x_{DM}$ is the charge of the
DM particle under the $U(1)_x$. Thus, for gauge boson masses of a
few GeV and values of $\epsilon$ towards the lower end of the
reasonable range, the DAMA signal can arise from IDM scattering
off nuclei mediated by a hidden $Z'$.

  Phenomenological bounds on a light hidden $U(1)_x$ have
been studied in Refs.~\cite{Appelquist:2002mw,Boehm:2003hm,
Feldman:2006ce,Chang:2006fp,Pospelov:2008zw}.
The strongest constraints on a hidden $U(1)_x$ gauge boson with
mass of a few GeV, coupling to the SM through kinetic mixing with
electromagnetism, comes  from measurements of the magnetic dipole
moments of the electron and muon. For masses larger than about a
GeV, values of the kinetic mixing parameter $\epsilon$ less than
$10^{-2}$ are acceptable~\cite{Pospelov:2008zw}.

  If the mass of the dark matter particle is much larger than the mass
of the light $U(1)_x$ gauge boson, a potential fermionic IDM candidate and its slightly
heavier partner will both annhilate very efficiently into gauge
boson pairs. The annihilation
cross section for this process is~\cite{Pospelov:2007mp}
\beq
\left<\sigma\,v\right>_{VV} \simeq
\frac{(g_x\,x_{DM})^4}{16\pi}\,\frac{1}{M_{DM}^2} \simeq
\lrf{g_x\,x_{DM}}{0.5}^4\lrf{500\,\gev}{M_{DM}}^2\, (5.8\times
10^{-26}\,cm^3/s).
\eeq
Note that this cross section scales as
$\sim g_x^4/M_{DM}^2$, just as for electroweak gauge boson
exchange. For reference, the necessary thermal DM relic density is
approximately $\Omega_{DM}\,h^2 \simeq 2\,(3\times
10^{-27}cm^3/s)/\left<\sigma\,v\right>$, where the additional
factor of two accounts for the fact that two different states are
annihilating~\cite{Gondolo:1990dk,Edsjo:1997bg}.
In addition, there are contributions to the annihilation cross sections
from the $s$-channel exchange of $U(1)_x$ gauge bosons decaying into
pairs of light matter fields in the $U(1)_x$ sector.

These contributions depend on the precise field content of the theory,
but they will typically be on the same order as for annihilations
into light gauge boson pairs. In general, the correct thermal
relic density of a fermionic IDM state in this scenario
will be obtained for a dark matter mass on the order of several
hundreds of GeV.  For a complex scalar IDM state, there is a
moderate additional velocity suppression of the annihilation rate,
and a somewhat lighter mass (but still on the order of a few hundred GeV)
will yield the correct thermal relic density~\cite{Boehm:2003hm}.

  It follows from the discussion above and Eq.~\eqref{lightu1sig}
that the we can choose parameters to give the correct thermal abundance
while simultaneously accounting for the DAMA signal.
Recall, however, that in the case of electroweak interactions
this came out automatically for the known gauge boson masses
and gauge coupling.

\section{Models of Inelastic Dark Matter\label{models}}

  Our guiding assumption in investigating candidates for IDM is that
the interaction between the DM state and the target nucleus is
mediated by a massive gauge boson.  Though it is technically natural for the mass splitting
operator coefficients to be small because they violate a global
$U(1)_{DM}$ symmetry,  the primary challenge is
to generate a small inelastic mass splitting in a reasonable way.
 With this in mind, we  introduce
a $U(1)_{DM}$-breaking spurion $\phi$.  Depending on the
relative charge of the dark matter field and $\phi$, the mass
splitting operator will be naturally suppressed by a factor
$(\frac{<\phi>}{\Lambda})^n$. This operator could be used
to generate a $\sim 100 \kev$ splitting for a judicious choice of
$n$ and $<\phi>$.

  This is not terribly satisfying, however, as it simply
parameterizes our ignorance about how the symmetry is broken.
We would also like to understand the physics underlying
the inelastic splitting and, given that the masses generally
hover around the electroweak scale,  fit it together with possible solutions
to the hierarchy problem.    The goal of this
section is to come up with simple models that can give rise to a
splitting of the correct size in the context of a solution to the
hierarchy problem without any large tuning of parameters. Of
course, in the end aesthetic criteria are subjective, so we view
this section as a compilation of interesting ways to generate
splittings of the right size for fermions or scalars.

\subsection{Models Mediated by the SM $Z^0$}

  As we saw in Section~\ref{general}, dark matter charged under
$SU(2)_L$ seems particularly promising because the $Z^0$-exchange
cross section is roughly the correct size to account for the DAMA
signal when one chooses a dark matter mass that gives the right
thermal relic abundance. This is especially interesting for
$SU(2)_L$ doublet dark matter, which is the simplest
possible representation.
However, the size of the mass splitting is then
somewhat of a mystery, with scalar dark matter requiring the
coefficient of the splitting operator Eq.~\eqref{scalarop} to be
$\lambda \sim 10^{-6}$, and fermion dark matter requiring the
splitting operator Eq.~\eqref{fermop} to be suppressed by the mass
scale $\Lambda \sim 10^8 \gev$.  Both splittings are smaller than
what one would na\"{\i}vely expect for an effective field theory
valid below the TeV scale if $U(1)_{DM}$ were strongly broken.

  One simple possibility is that the effects of $U(1)_{DM}$ breaking
are sequestered in some way.  This could happen, for example, if
the splitting operators are generated by integrating out a singlet
$S$ through an operator $D S^* h$, where $h$ is the SM Higgs field
and $D$ is the dark matter doublet.
In this case, $U(1)_{DM}$ breaking could be communicated through
the singlet. If the singlet is very heavy, or only couples weakly
to the doublet $D$, the mass splitting is suppressed.
Note that if there are multiple trilinear couplings $D S^* h$
and $D S h$, we can simply call the linear combination that
couples to the doublet $S^*$ and define the $U(1)_{DM}$
symmetry so as to respect this trilinear coupling.

  There are then two ways that $U(1)_{DM}$ breaking could be
communicated. The first possibility is that most of the singlet
mass is $U(1)_{DM}$ preserving, with a small $U(1)_{DM}$ breaking
piece.  For a scalar with the potential
\beq
\mathscr{L} \supset - m_S^2 |S|^2 - \left(\frac{m_{\delta}^2}{2}
S^2 + f D S^* h + h.c.\right),
\eeq
integrating out the
singlet generates the operator $\frac{f^2
m_{\delta}^2}{m_S^4} D D h h$.  For a fermion with Lagrangian
\beq
\mathscr{L} \supset - m_S \bar{S} S -
\left(\frac{m_{\delta}}{2} \bar{S}^c S + \lambda \bar{D} S h +
h.c.\right),
\eeq
 integrating out the singlet  generates
the operator $\frac{\lambda^2 m_{\delta}}{m_S^2}\bar{D}^c D h h$.
If there is a natural hierarchy between $m_{\delta}$ and $m_S$, or
if the couplings $\lambda$ or $f$ are naturally small, one could
obtain a splitting of the correct size.

  The second possibility is that the singlet has a very large
$U(1)_{DM}$ breaking mass $m_{\delta}$.
This would, for example, be the only option if the singlet is a real scalar
or a Majorana fermion.
Integrating out the singlet would then generate the scalar operator
$\frac{f^2}{m_{\delta}^2} D D h h$ or the fermion operator
$\frac{\lambda^2}{m_{\delta}} \bar{D}^c D h h$. This is
completely analogous to the way small neutrino masses are
induced in the conventional seesaw mechanism.\footnote{In fact,
it may be possible that the same right-handed neutrino scale
enters both the neutrino mass and the dark matter splitting
operators -- this would require that the lepton doublets come with
an additional suppression factor, but this may be natural in a
model of flavor physics.} A splitting of the right size could
again be obtained for small couplings or if $m_S$ is naturally
identified with an intermediate scale.

  Only a handful of concrete models exist for IDM whose nucleon scattering
is mediated by the $Z^0$. Ref.~\cite{TuckerSmith:2001hy} 
proposed a model of left-handed sneutrino dark matter 
in which the inelastic mass splitting is generated through mixing with a scalar
singlet right-handed sneutrino. The $U(1)_{DM}$ violating (and
lepton-number violating) mass arises through a SUSY breaking
operator $\frac{1}{M_{Pl}^3} X^{\dagger} X^{\dagger} X N^{\dagger}
N$, and the size of the splitting is naturally related to an
intermediate scale. For related models, 
see Refs.~\cite{Arina:2008bb,Kumar:2008vs}.
A second possibility for IDM are the Dirac
neutralinos that arise in $U(1)_R$-symmetric SUSY
scenarios~\cite{rsymidm1,rsymidm2}. The small Majorana mass
splitting would then be related to a small amount of $U(1)_R$
breaking.

  We present below several other models for $SU(2)_L$ doublet
IDM that make use of a singlet to communicate $U(1)_{DM}$ breaking.
Two of these models are based on a warped extra
dimension~\cite{Randall:1999ee}, and illustrate some of the
ways an inelastic splitting can emerge in this context.  We also
present a supersymmetric model.

\subsubsection{Warped Fermion Model\label{seesaw}}

  We begin with a model of fermion $SU(2)_L$ doublet dark matter
and attempt to explain how the scale of splitting $\delta \sim 100
\kev$ can emerge without any large hierarchy of input parameters.
This requires  an explanation for the scale suppressing the
splitting operator Eq.~\eqref{fermop}, $\Lambda \sim 10^8 \gev$,
which could represent the mass of a singlet field  that has been integrated out.
One way that this intermediate scale mass could
emerge naturally is if it is equal to the Planck scale times an
exponential suppression factor. As we will see below, it is
straightforward to realize this possibility in the context of a 5D
Randall-Sundrum model~\cite{Randall:1999ee}, giving the added
bonus of combining a natural dark matter model with a solution to
the hierarchy problem. Our model is similar to the models of
Refs.~\cite{Huber:2003sf,Perez:2008ee},
which realize the seesaw mechanism in warped geometry.

  In particular, we consider $AdS_5$ compactified on $S_1 / \mathbb{Z}_2$
with metric \beq \label{AdS} ds^2 = e^{- 2 k |y|} \eta_{\mu \nu} d
x^{\mu} d x^{\nu} - d y^2, \eeq where $ - \pi R \leq y \leq \pi
R$. If the dark matter derives from a vector-like $SU(2)_L$
fermion doublet $D = (D_L,D_R)^T$ localized on the TeV brane, its
mass is naturally of order the TeV scale. However, since we also
expect the Higgs doublet to be localized to the TeV brane, we need
to forbid the TeV brane localized operator
$\frac{1}{\Lambda_{\tev}} \bar{D}^c D h h$ which would generate
too large of a splitting.  We therefore impose a $U(1)_{DM}$
symmetry under which $D$ is charged to forbid the splitting
operator, and assume that this symmetry is broken only on the UV
brane.

  In order to communicate the breaking of $U(1)_{DM}$,
we introduce a bulk fermion singlet $S = (S_L,S_R)^T$ with $(+,+)$
boundary conditions for $S_L$.  We include a $U(1)_{DM}$-breaking
Majorana mass on the UV brane, along with a $U(1)_{DM}$-preserving
bulk mass.  The singlet action is taken to be
\begin{eqnarray}
S &=& \int d^4 x \int d y \sqrt{-g} \left[ i \bar{S} \gamma^M D_M
S + c k \epsilon(y) \bar{S} S - \delta(y) \left( \frac{d_{UV}}{2}
\bar{S}^c_L S_L + h.c. \right)
\right. \nonumber \\
&& \left.\phantom{\frac{d_{UV}}{2}} - \delta(y - \pi R)
\left(\lambda \bar{D}_R S_L h + h.c.\phantom{\frac{.}{.}}\!\!\!
\right) \right]
\end{eqnarray}
where only the left-handed component of the singlet can have brane
couplings because of the choice of boundary conditions.  Here,
$S^c = C \gamma^0 S^*$ where $C$ is the 5D charge conjugation
operator.  The sign of the bulk mass parameter $c$ has been chosen
to agree with the convention that $c > 1/2$ localizes a zero mode
towards the UV brane.  Note that while the boundary singlet mass
explicitly violates the $U(1)_{DM}$ symmetry, there remains an
unbroken $\mathbb{Z}_2$ subgroup under which $D$ and $S$ are odd
ensuring that the lightest of these fermions is stable.

  To see the effect of the boundary terms on the fermion masses,
it is easiest to first expand the singlet in a basis that diagonalizes
the KK modes \emph{without} the boundary terms.  In this basis,
the communication of $U(1)_{DM}$ breaking is dominated by the chiral
zero mode which picks up a large Majorana mass.  Since the remaining
modes acquire Dirac KK mass terms, we can truncate the KK tower
while still capturing the dominant contribution.  In particular,
one can expand
\beq
S_{{ }^L_R} (x^{\mu}, y) = \frac{e^{2 k
|y|}}{\sqrt{2 \pi R}} \sum_{n=0}^{\infty} S_{{ }^L_R}^{n}
(x^{\mu}) f_{{}^L_R,n} (y),
\eeq
where the wavefunctions $f_{{}^L_R,n}$ solve the bulk
equations of motion
\beq (\partial_y
\pm c k) f_{{ }^L_R,n}(y) &=& \mp m_n e^{k |y|} f_{{ }^R_L,n}(y).
\eeq

  Imposing $(+,+)$ boundary conditions for $S_L$, the solutions are
~\cite{Gherghetta:2000qt}
\beq
f_{{ }^L_R,n}(y) =
\frac{e^{k |y|/2}}{N_n} \left [ J_{-c \mp \frac12} (\frac{m_n}{k}
e^{k |y|} ) - \frac{J_{-c + \frac12} (\frac{m_n}{k} e^{\pi k
R})}{Y_{-c +\frac12} (\frac{m_n}{k} e^{\pi k R} )} Y_{-c \mp
\frac12} (\frac{m_n}{k} e^{k |y|}) \right ] .
\eeq
where the
masses $m_n$ can be determined from the condition that $f_{R,n}(0)
= 0$, and the normalization factors are obtained from
\beq
\frac{1}{2 \pi R} \int_{-\pi R}^{\pi R} d y\; e^{k |y|}
\left(f^*_{L,m} f_{L,n} + f^*_{R,m} f_{R,n} \right) = \delta_{m n}
.
\eeq
In particular, there is a massless chiral zero mode with
\beq
f_{L,0}(y) = \sqrt{\frac{(2 c - 1) \pi k R}{1 - e^{- (2 c -
1) \pi k R}}} e^{-c k |y|}.
\eeq
The other KK masses and
normalization factors can be approximated for $m_n << k$ and $k R
>> 1$ as~\cite{Gherghetta:2000qt}
\beq
m_n \approx \left( n - \frac{c}{2} \right) \pi k e^{- \pi k R}
\eeq
assuming $c < 1/2$, and
\beq
N_n \approx \sqrt{\frac{2}{\pi^2\, R\, m_n}} e^{\pi k R /2}.
\eeq

  After performing this decomposition, the 4D fermion mass matrix is
not yet diagonal due to the brane localized mass terms.  The
singlet mass matrix is
\beq
\mathscr{L} \supset - \frac12 \left(
\begin{array}{cccc} \bar{S}^0_L & \bar{S}^1_L & \bar{S}^{1 c}_R & ...
\end{array} \right) \left( \begin{array}{cccc}
  A_{0 0} & A_{0 1} & 0 & ... \\
  A_{0 1} & A_{1 1} & m_1 & ... \\
  0 & m_1 & 0 & ... \\
  \begin{sideways}...\end{sideways} & \begin{sideways}...\end{sideways} &
\begin{sideways}...\end{sideways} & {}
\end{array}
\right)
\left( \begin{array}{c} S_L^{0 c} \\ S_L^{1 c} \\ S_R^{1} \\
\begin{sideways}...\end{sideways} \end{array} \right) + h.c.
\eeq
where the Majorana masses $A_{m n}$ are given in terms of the
wavefunctions on the UV brane
\beq
A_{m n} \equiv \frac{d_{UV}}{2
\pi R} f_{L,m}(0) f_{L,n}(0).
\eeq
In particular, the zero mode
picks up a mass
\beq A_{0 0} = \frac{d_{UV} k (c - 1/2)}{1 - e^{-(2c - 1) \pi k R}}.
\eeq

  The couplings to the canonically normalized doublet and Higgs
field on the TeV brane are \beq \mathscr{L} \supset -
\sum_{n=0}^{\infty} C_n \bar{D}_R S_L^{n} h + h.c. \eeq where \beq
C_n \equiv \frac{\lambda e^{\pi k R / 2}}{\sqrt{2 \pi R}}
f_{L,n}(\pi R) \eeq is determined from the wavefunction overlap on
the TeV brane. Assuming that we choose c such that $A_{0 0} >>
m_n$, we can to a good approximation simply integrate out the
heavy chiral $S_L^0$ mode. This generates the splitting operator
\bea \mathscr{L} &\supset&
\frac{C_0^2}{2 A_{0 0}} \bar{D}^c_R D_R h h + h.c. \\
 &=& \frac{\lambda^2}{2 d_{UV}} e^{- 2 (c - 1/2) \pi k R}
\bar{D}^c_R D_R h h + h.c.
\eea
Choosing the natural values
$\lambda^2 \sim \frac{1}{M_{pl}}$ and $d_{UV} \sim 2$, we need $c
\sim 0.13$ to obtain a mass splitting on the order of $\delta \sim
100 \kev$. It is straightforward to check that including the KK
modes gives a negligible ($\sim 10^{-4}$) correction to these
estimates (see Appendix~\ref{appa}). This is due to the fact that,
unlike the former zero mode, the states in the KK tower are mostly
Dirac and are very poor at communicating $U(1)_{DM}$ breaking.

  Arranging the model parameters to produce a mass splitting
on the order of $\delta \sim 100\kev$, our model has all the
ingredients needed to account for the DAMA annual modulation
result. The lightest stable fermion state is an almost pure
$SU(2)_L$ doublet Majorana fermion with a slightly heaver
inelastic partner, and is stable on account of the unbroken
$\mathbb{Z}_2$ subgroup of $U(1)_{DM}$. If the corresponding
fermion mass is $\sim 1.1 \tev$, it will the yield the correct
relic density from thermal freeze-out, in addition to the correct
nucleon scattering cross section and mass splitting as was shown
in Section~\ref{general}.

\subsubsection{Warped Scalar Model \label{scalarmodel}}

We now turn to a model of scalar $SU(2)_L$ doublet dark matter. To
sequester $U(1)_{DM}$ breaking we again consider a warped 5D setup
with the metric of Eq.~\eqref{AdS}, and assume that a complex
scalar doublet $D = \frac{1}{\sqrt{2}}(D_R + i D_I)$ is localized
to the TeV brane in addition to the Higgs field.  As in the
previous model, we need to forbid the TeV brane-localized operator
$D D h h$, which would split the masses of $D_R$ and $D_I$ by
$\sim \frac{v^2}{m_D}$ with an $O(1)$ coupling.  We again  assume a
$U(1)_{DM}$ symmetry which is broken only on the UV
brane, while preserving a $\mathbb{Z}_2$ subgroup to ensure the stability
of the doublet DM.

In order to communicate this $U(1)_{DM}$ symmetry breaking to the
DM doublet, we proceed as in the fermion model and introduce a
bulk (scalar) singlet.  Note that we cannot allow the doublet to
directly couple to the UV symmetry breaking, even though a
wavefunction localized to the TeV brane can be sufficiently
suppressed in the UV,  because the UV mass would then break
hypercharge (since a neutral dark matter
doublet must carry hypercharge) at too high a
scale.  Instead we assume the IR-brane-localized doublet mixes
with a bulk singlet, as in the fermion model of the previous
section.

  The singlet can then couple in symmetry-breaking operators on the
UV brane and communicate this breaking to the doublet on the TeV
brane.  However, we note a critical difference between 5D scalars
and fermions -- in general bulk scalars do not possess a zero
mode. This means that the only option is to communicate
$U(1)_{DM}$ breaking through singlet KK modes, which naturally
peak away from the UV and tend to pick up small
$U(1)_{DM}$-breaking masses (in addition to their
$U(1)_{DM}$-preserving masses).  Therefore this model does not
suppress mass splitting via a seesaw mechanism.  Instead, the real
and imaginary components of the doublet will be split by their
mixing to the split components of the bulk scalar KK modes.

  In fact, bulk KK modes suppress the communication of $U(1)_{DM}$
symmetry breaking to the doublet extremely effectively -- so much
so that the effective UV scale cannot be much higher than $\sim
100-1000\,\tev$ or the $U(1)_{DM}$ breaking would effectively
decouple and give too small a mass splitting to the dark matter
candidate.  In the context of an RS solution to the
hierarchy problem, this requires a setup with either a third
brane,\footnote{Detailed considerations of the three-brane setup,
such as ensuring its stability, are beyond the scope of this work.
Related references can be found in, \emph{e.g.},
Refs.~\cite{Kogan:2000xc,Moreau:2004qe}.} or an additional warped
dimension such as was considered in Ref.~\cite{McDonald:2008ss}.
Alternatively, one can simply treat the smaller warp factor
as giving a solution to the flavor hierarchy problem, such as in
the little RS scenario of Refs.~\cite{Davoudiasl:2008hx,Davoudiasl:2008nr}.
The model of this section can be thought of as a two-brane effective
theory describing any of these situations, generated by integrating
out physics above the effective UV scale.

We thus start by introducing a complex bulk singlet $S$ with
$(+,+)$ (Neumann) boundary conditions.  In general, $S$ has a
bulk mass and two $U(1)_{DM}$-preserving brane mass terms, as well
as a $U(1)_{DM}$-breaking mass term on the UV brane.  To simplify
the calculation we assume that the $U(1)_{DM}$-preserving brane
mass terms are negligible.  We consider the singlet action
\bea
S &=& \int d^4x \int dy\sqrt{-g}
\left[ \partial_M S^*\partial^M S - m^2 S^*S
- \delta(y)\left(\frac{m_{UV}}{2}\,S^2+h.c.\right) \right.
\nonumber \\
&& \left. \phantom{\frac{m_{UV}}{2}} - \delta(y - \pi R) \left(
\lambda\, e^{2 \pi k R}\, D S^* h + h.c. \right) \right],
\label{singletscalar}
\eea
where $m \equiv \sqrt{a} k$ is the
$U(1)_{DM}$-preserving bulk mass of $S$, $m_{UV}$ is the
$U(1)_{DM}$-violating UV brane mass, and $\lambda$ is the coupling
to the dark matter field $D$.  The exponential factor is because
$D$ and $h$ are taken to be canonically normalized 4D fields,
so $\lambda \sim \sqrt{k}$ has units of $\sqrt{\text{mass}}$.

  The KK decomposition of $S$ is
\beq
S(x^{\mu},y) =
\frac{1}{\sqrt{2 \pi R}} \sum_n S^{n}(x^{\mu}) f_n(y),
\eeq
where the functions $f_n(y)$ that solve the bulk equations of motion
are~\cite{Gherghetta:2000qt}
 \beq
f_{n}(y) =
\frac{e^{2 k |y|}}{N_n} \left[ J_{\alpha} (\frac{m_n}{k} e^{k
|y|}) + b_{\alpha} (m_n) Y_{\alpha} (\frac{m_n}{k} e^{k |y|})
\right]. \label{scalarwavefunction}
\eeq
Here, $\alpha = \sqrt{4+a}$, and the normalization factor $N_n$ can be
approximated in the limit $m_n << k$ and $k R >> 1$ as
\beq N_n
\approx \sqrt{\frac{1}{\pi^2\, R\, m_n}} e^{\pi k R / 2}.
\eeq
The masses $m_n$ and the functions $b_{\alpha}$ are determined
from the boundary conditions.  In the $U(1)_{DM}$-symmetric limit
($m_{UV}=0$), the boundary conditions of $S_R$ and $S_I$ are
identical, so there is a pair of degenerate states at each KK
level.  When $U(1)_{DM}$ is broken ($m_{UV} \neq 0$), $S_R$ and
$S_I$ have different boundary conditions and the resulting KK
masses are split.

 In the presence of the $U(1)_{DM}$-violating mass, variation of
the boundary action determines the boundary conditions to be
\bea
    \partial_y S_{{ }^R_I} \mp m_{UV} S_{{ }^R_I} &=& 0 \vert_{y=0}\\
\nonumber
    \partial_y S_{{ }^R_I} &=& 0 \vert_{y = \pi R}.
    \label{scalarbc}
\eea
It is straightforward to solve numerically for the KK mode
mass splitting in the presence of these boundary conditions.
This mass splitting in the singlet sector is then communicated to the
DM doublet through the couplings on the TeV brane.

  It is easier however to proceed as in the last section and
first expand the KK states in the basis without the splitting,
treating the $U(1)_{DM}$-violating mass as a perturbation.
 Imposing the simpler boundary conditions $\partial_y S = 0
\vert_{y = 0, \pi R}$ leads to the two conditions
\bea
b_{\alpha}(m_n)
&=& - \frac{ 2 J_{\alpha} (\frac{m_n}{k})
+ \frac{m_n}{k} J'_{\alpha} (\frac{m_n}{k}) }{ 2 Y_{\alpha} (\frac{m_n}{k})
+ \frac{m_n}{k} Y'_{\alpha} (\frac{m_n}{k}) } \\
&=& - \frac{ 2 J_{\alpha} (\frac{m_n}{k} e^{\pi k R}) +
\frac{m_n}{k} e^{\pi k R} J'_{\alpha} (\frac{m_n}{k} e^{\pi k R})
}{2 Y_{\alpha} (\frac{m_n}{k} e^{\pi k R}) + \frac{m_n}{k} e^{\pi
k R} Y'_{\alpha} (\frac{m_n}{k} e^{\pi k R}) }
\eea
which yield the approximate KK spectrum in the limits $m_n << k$
and $k R >> 1$~\cite{Gherghetta:2000qt},
\beq
m_n \approx \left(n + \frac{\alpha}{2}
- \frac34\right)\,\pi\,k\,e^{- \pi k R} .
\eeq
In this basis, the masses in
the singlet sector can be written as
\bea
\mathscr{L}&\supset&
-\frac{1}{2} \left(\begin{array}{ccc}
S^1_R & S^2_R & ... \\
\end{array}\right)
\left(
\begin{array}{ccccc}
m_1^2 + \Delta_{1 1}^2 & \Delta_{1 2}^2 & ... \\
\Delta_{2 1}^2 & m_2^2 + \Delta_{2 2}^2 & ... \\
\begin{sideways}...\end{sideways} & \begin{sideways}...\end{sideways} & {}\\
\end{array}
\right)\left(
\begin{array}{c}
S^1_R\\
S^2_R \\
\begin{sideways}...\end{sideways} \\
\end{array}
\right) \nonumber\\
&&~~~ - \frac{1}{2} \left(\begin{array}{ccc}
S^1_I & S^2_I & ... \\
\end{array}\right)
\left(
\begin{array}{ccccc}
m_1^2 - \Delta_{1 1}^2 & - \Delta_{1 2}^2 & ... \\
- \Delta_{2 1}^2 & m_2^2 - \Delta_{2 2}^2 & ... \\
\begin{sideways}...\end{sideways} & \begin{sideways}...\end{sideways} & {}\\
\end{array}
\right)\left(
\begin{array}{c}
S^1_I\\
S^2_I \\
\begin{sideways}...\end{sideways} \\
\end{array}
\right) \eea
where
\beq
\Delta^2_{m n} \equiv \frac{m_{UV}}{2 \pi R} f_m(0) f_n(0).
\eeq

  The mass splitting in the singlet sector at each KK level is then
determined by the difference in the eigenvalues of these two mass
matrices.  As long as we choose parameters such that $\Delta_{m n}
<< m_n$, the mixing is small and the eigenvalues are simply $m_n^2
\pm \Delta_{n n}^2$ up to $O(\frac{\Delta^4}{m^2})$ corrections.
Expanding the wavefunction profiles in
Eq.~\eqref{scalarwavefunction} for $m_n << k$ at $y = 0$, the mass
splitting at the $\text{n}^{\text{th}}$ KK level  is then
approximately
 \bea
\Delta m_n &\approx& \frac{\Delta_{n n}^2}{m_n} \nonumber\\
&\approx&\left(\frac{m_n}{k}\right)^{2 \alpha}
\left[\frac{\pi\,e^{- \pi k R}\,m_{UV}\,}{2^{2\alpha - 1}\,
\Gamma(\alpha)^2\, (\alpha - 2)^2}\right]. \label{mnsplitting}
\eea

The mass splitting in the singlet sector is transmitted to
the DM doublet through their coupling on the TeV brane.  After the
Higgs acquires a VEV, the mass matrix is
 \beq
\mathscr{L}\supset -\frac{1}{2}\!\left(\begin{array}{ccccc}
D_R&D_I&S^1_R&S^1_I&...\\
\end{array}\right)\!\!
\left(
\begin{array}{ccccc}
m_D^2&0&C_1 v&0&...\\
0&m_D^2&0&C_1 v&...\\
C_1 v&0&m_1^2+\Delta_{1 1}^2&0&...\\
0&C_1 v&0&m_1^2-\Delta_{1 1}^2&...\\
\begin{sideways}...\end{sideways}&\begin{sideways}...\end{sideways}&\begin{sideways}...\end{sideways}&\begin{sideways}...\end{sideways}&{}\\
\end{array}
\right)\!\!\!\left(
\begin{array}{c}
D_R\\
D_I\\
S^1_R\\
S^1_I\\
\begin{sideways}...\end{sideways} \\
\end{array}
\right) \eeq where \beq C_n \equiv \frac{\lambda e^{-2 \pi k
R}}{\sqrt{2 \pi R}} f_n(\pi R) .
\eeq

It is straightforward to diagonalize this mass matrix at a given
level of truncation of the KK tower.  For example, including just
the first KK state leads to a mass squared splitting to leading
order in $\Delta_{1 1}$ of \beq \Delta m_D^2 \approx \Delta_{1
1}^2 \left[1 - \frac{|m_1^2-m_D^2|}{\sqrt{(m_1^2-m_D^2)^2 + 4
C_1^2 v^2}} \right]. \eeq Note that this goes to zero if either
the UV brane mass or TeV brane coupling turns off, just as one
would expect. We can further expand this for small $v << m_n$, and
write the contribution from the $\text{n}^{\text{th}}$ KK level
more generally as \beq \Delta m_D \approx \Delta m_n \frac{C_n^2
v^2}{m_D m_n^3} \left(1+\frac{2 m_D^2}{m_n^2}\right),
\label{mdsplit} \eeq where $\Delta m_n$ is the singlet mass
splitting at the nth KK level, given approximately in
Eq.~\eqref{mnsplitting}.  Choosing the values $k \sim 500\tev$, $R
\sim 2.1 / k$, $m_{UV} \sim 2 k$, $a \sim 0.1$, $\lambda \sim 2.5
\sqrt{k}$, and $m_D \sim 525\gev$, for example, gives a
contribution to the DM mass splitting from the first KK mode of
$\sim 5\kev$.\footnote{Note that while $k R \sim 2.1$ is not much
larger than 1 as was assumed in some of the approximations above,
we find that the analytical formulae still give a reasonable
approximation to the full numerical results.} Note that a
splitting in the singlet sector that is not too small requires the
warp factor $kR$ not be too large.

However, when trying to sum these KK contributions, the sum
apparently diverges.  This is counter-intuitive, since we would
expect heavy modes to decouple.  Here we might expect this
decoupling to happen since the contribution to the splitting
appears to be smaller for heavier KK modes due to the $\sim 1/n^3$
suppression in Eq.~\eqref{mdsplit}.  However, this is compensated
for by the fact that the singlet splitting $\Delta m_n$ is
increasing for higher KK modes due to a larger UV brane overlap,
and scales roughly as $\sim n^{2 \alpha} \gtrsim n^4$.

  The issue then is the behavior of $C_n$ for higher KK modes.  In
the na\"{\i}ve limit of an infinitely thin brane, the higher KK
modes would appear to couple to the TeV brane with roughly equal
strength, and the magnitude of $C_n$ would not significantly
change as one increases $n$.  This behavior was seen, \emph{e.g.},
in Refs.~\cite{Hewett:2002fe,Pomarol:1999ad}, where the IR brane
coupling of KK SM gauge fields was studied and found to be
universal.  This would lead to a contribution to the DM mass
splitting that increases approximately linearly with $n$ for small
values of the bulk mass $a$.

However, this result is unphysical since it does not take into
account the thickness of the IR brane, $\Delta \sim \Lambda^{-1}$,
where $\Lambda$ is the 5D cutoff scale and the thickness is given
in $y$-coordinate space.  We expect $\Lambda\sim (10\!-\!100)\, k$
based on na\"ive dimensional analysis.
Note that the physical thickness of the brane at $y \simeq \pi R$
is redshifted to $\Delta_{phys} \sim \Lambda^{-1} e^{\pi k R}$.
The effective 4D coupling $C_n^{eff}$ is then better approximated
by integrating the wavefunctions over this thickness. While not
terribly important for lower KK modes, higher KK modes rapidly
oscillate over this region and the effective coupling is
suppressed due to a cancelation between opposite phases.

  While the exact numerics depend on the details of the Higgs and DM
doublet wavefunction profiles over the brane thickness, we can
understand this decoupling by assuming the ``flat'' profiles \beq
f_{h,D}(y) \simeq \frac{1}{\sqrt{\Delta}}e^{k y}\label{5dHD}, \eeq
and concretely taking $\Delta^{-1} \sim 10 \,k$. The effective
coupling $C_n^{eff}$ is then given by \beq C_n^{eff} \simeq
\frac{\lambda}{\sqrt{2 \pi R}} \frac{1}{\Delta} \int_{\pi R -
\frac{\Delta}{2}}^{\pi R + \frac{\Delta}{2}} d y \,e^{-2 k y}
\,f_n(y). \label{cneffeq} \eeq With these choices, only the first
$\sim \Lambda / k \sim 10$ KK modes   make a significant
contribution to the DM mass splitting.  Taking the same parameters
as before, we obtain a mass splitting of $\sim 70\kev$ when summing
over the first 10 KK modes.  Note that this decoupling behavior
itself is quite robust in that it will happen for any reasonable
choice of Higgs and DM doublet profiles.

For higher values of $n$, the integrand in Eq.~\eqref{cneffeq} is
rapidly oscillating over the brane thickness $\Delta$.  This
happens when the KK modes have enough 5D momentum such that many
wavelengths $\lambda_n$ fit inside the physical thickness of the
brane. We then expect  a phase cancelation up to terms suppressed
by about $\sim \lambda_n / \Delta_{phys} \sim 1/n$ (see
Appendix~\ref{appb} for a more detailed derivation).  For small
bulk mass $a$, $\Delta m_n$ in Eq.~\eqref{mnsplitting} scales
roughly as $\sim n^4$, so the mass splitting in
Eq.~\eqref{mdsplit} now scales as $\sim 1/n$. Summing the
contributions from these higher KK modes up to $n \simeq
\frac{\Lambda}{k} e^{\pi k R}$ (corresponding to modes with
momentum around the 5D cutoff scale) then gives an additional
contribution which is enhanced by $\sim \log(e^{\pi k R}) \sim 6$
relative to the contribution from a single lower mode. Estimating
the overall amplitude as $\sim 1/10$ of the sum of the first 10
modes, these higher KK modes can then give an $O(1)$ correction to
the mass splitting, and one can obtain a splitting of the desired
size.

\subsubsection{A Supersymmetric Candidate}

  The mechanism described above for generating a splitting, namely mixing the
DM doublet with a singlet that has a $U(1)_{DM}$-breaking mass,
can apply in other contexts as well.  We next consider the
application of this mechanism in the context of low-energy
supersymmetry.  In this case, a natural small
suppression can arise for example in large
$\tan\beta$ scenarios when mixing the singlet with the doublet dark matter
candidate through the VEV of $H_d$.

An example of a  model that exploits this suppression consists of the MSSM augmented by a vector pair of
$SU(2)_L$ doublet chiral superfields $D$ and $D^c$ with $Y=\pm
1/2$.  Hypercharge and holomorphy then allow the superpotential
operator $DD^c$, but not $DD$ or $D^cD^c$, and an accidental
$U(1)_{DM}$ global symmetry can arise.  This
symmetry must be broken to generate an inelastic splitting.  To do so, we gauge a
related $U(1)_z$ symmetry and break it by Higgsing.  A simple
superpotential that  realizes the above symmetries is
\beq
W \supset
\lambda\,N\,H_u\ccdot H_d + \lambda'\,S\,H_d\ccdot D
+\frac{\xi}{2}\,N\,S^2 + \zeta\,N\,DD^c. \label{w1}
\eeq
Here, $N$ and $S$ are SM singlets that carry non-zero charges under the
gauged $U(1)_z$ group.

If the $N$ field develops a VEV induced by supersymmetry breaking
soft terms, $N \to \left<N\right> \sim \tev$, a supersymmetric
mass for $S$ will be generated.\footnote{The size of the $N$ VEV
can be set by SUSY breaking so that it is naturally on the order
of a TeV. The $N$ scalar will get a mass upon expanding the scalar
potential, while the $N$ fermion will develop a mass by mixing
with the $U(1)_z$ gaugino.}  To get an approximate picture of what
this does to the masses of the fermions in the model, we can
integrate out the $S$ superfield in the supersymmetric limit. This
yields
\beq
W_{eff} \supset \lambda\,\left<N\right>\,H_u\ccdot H_d
+ \zeta\,\left<N\right>\,DD^c
-\frac{{\lambda'}^2}{2\xi\left<N\right>}\,(H_d\ccdot D)^2.
\label{weff1}
\eeq
The last term is the desired mass splitting
operator.  It receives a suppression from large $\tan\beta$, our
choice that $\lambda' < 1$, and from an assumed small hierarchy between
$\left<N\right>$ and the electroweak scale.  A more careful
analysis of mixing in the SM-neutral fermion sector after symmetry
breaking shows that the fermion mass eigenstates consist of an
almost pure singlet and a nearly degenerate pair of Majorana
states that derive almost entirely from the doublets.  The
resulting SM-neutral fermion masses are
\beq
M_{d_{{}^1_2}} =
\zeta\,\left<N\right> + \delta_{\pm}, ~~~~M_s =
\xi\,\left<N\right> + \delta_{+} - \delta_{-},
\label{fermmass}
\eeq
with
\beq
\delta_{\pm} \simeq \pm \frac{{\lambda'}^2v_d^2}{2M_s}
\left(1 \pm \frac{M_d}{M_s}\right)^{-1}.
\label{delferm}
\eeq
 Taking $\tan\beta = 30$,\; $M_s = 3000\,\gev$,\; $M_d = 1000\,\gev$,
and $\lambda' = 0.1$, we obtain a mass splitting of $\delta_{+} -
\delta_{-} \simeq 130\,\kev$.

  The neutral fermion components of $D$ and $D^c$ can therefore
yield IDM provided these fields are stable.  An unbroken
$\mathbb{Z}_2$ discrete symmetry is the minimal possibility to
ensure this.  If this symmetry is $R$-parity, $D$ and $D^c$ as
well as the $N$ and $S$ superfields must all be even.  On the
other hand, with a new $\mathbb{Z}_2$ it is possible for $D$,
$D^c$, and $S$ to be odd, with $N$ and the Higgs fields even.  The
inclusion of such a new discrete symmetry is well-motivated in
gauge-mediated models with a light gravitino.

  The form of the superpotential in Eq.~\eqref{w1} can be enforced
by the $U(1)_z$ charges
\bea \label{charges}
[S]_z = z_s,~~[N]_z &=& -2z_s,~~[D]_z = z_d,~~[D^c]_z= 2z_s-z_d,\\
\phantom{.}[H_d]_z &=& -z_s-z_d,~~[H_u]_z = 3z_s+z_d.\nnmb
\eea
 With these charges, the dangerous operator $[H_u\ccdot d^c]_z =
5z_s$ is forbidden provided $z_s\neq 0$, as is the bare $\mu$ term
operator $[H_u\ccdot H_d]_z = 2z_s$.  Note also that if $z_d=z_s$,
a $\mathbb{Z}_2$ that stabilizes both $D$ and $D^c$ arises
automatically as an unbroken discrete subgroup of the $U(1)_z$
gauge symmetry.  As it stands, this theory has mixed
$SU(2)_L^2\times U(1)_z$, $U(1)_Y^2\times U(1)_z$, and
$U(1)_z^2\times U(1)_Y$ anomalies when $z_s\neq 0$, implying that
SM-charged exotics must also be present in the theory.  In fact,
such exotics are necessary in any MSSM gauge extension that
forbids a bare $\mu$-term~\cite{Erler:2000wu,Morrissey:2005uz}.
 These exotics need not interfere with the dynamics discussed here.

  It is more challenging to obtain an acceptable scalar inelastic
dark matter splitting from this model.  The scalar components of
$D$ and $D^c$ will be stable if these superfields (and $S$) are
odd under $R$-parity or if we impose a new $\mathbb{Z}_2$
discrete symmetry.  In the latter case, it is necessary to tune
the soft masses such that the fermion components of $D$ and $D^c$
are heavier than the scalars.  At the supersymmetric level, a
scalar mass splitting arises from the $F$-term potential for $H_d$
derived from the effective superpotential of Eq.~\eqref{weff1}
\bea \label{delscal1}
V_F &\supset& \left|
\frac{\lambda'^2}{\xi\,\left<N\right>}\,H_d\ccdot
\tilde{D}\,\tilde{D}
+\mu_{eff}\,H_u\right|^2 \\
&\supset&
M_d\,\lrf{\lambda'^2\mu_{eff}\,v_uv_d}{M_d\,M_s}\,\tilde{D}\tilde{D}
+ h.c., \nnmb \eea with $M_d$ and $M_s$ as in
Eq.~\eqref{fermmass}, and $\mu_{eff} = \lambda\,\left<N\right>$.
Relative to the fermion splitting of Eq.~\eqref{delferm}, this
operator is suppressed by only a single power of $\cos\beta \sim
1/\tan\beta$ at large $\tan\beta$.  If generic soft supersymmetry
breaking operators are also included, there arises a further
scalar mass splitting with no $\cos\beta$ suppression at all. This
contribution can be thought of as coming from an operator of the
form \beq \label{delscal2} \mathscr{L}_{eff} \supset -
M_d\,\lrf{\lambda'^2A_{\xi}\,\mu_{eff}^2v_u^2}
{M_d\,M_s^3}\,\tilde{D}\tilde{D} + h.c., \eeq where $A_{\xi}$ is
the trilinear soft parameter corresponding to the $\xi\,N\,S^2$
superpotential operator.  Numerically, we find that the operators
of Eq.~\eqref{delscal1} and \eqref{delscal2} produce too large of
an inelastic scalar mass splitting unless $\lambda' \lesssim 0.01$
and there is an additional small hierarchy between
$\mu_{eff},\,A_{\xi}$ and $M_d,\,M_s$.  Note that even when the
scalar splitting is too large, the radiative corrections to the
fermion mass splitting are still safely small.

Scattering off nuclei by the IDM candidates that arise in this
model will be mediated primarily by the SM $Z^0$.  The massive
$U(1)_z$ gauge boson can also contribute to nuclear scattering,
but the effect will be suppressed by its larger mass.  If the new
states in the $U(1)_z$ sector are somewhat heavier than the IDM
candidate, the thermal relic density will be determined primarily
by electroweak interactions, and our estimates from
Section~\ref{general} for the relic density carry through.  An
interesting additional possibility arises when the doublet IDM
state is stabilized by a new $\mathbb{Z}_2$ symmetry, rather than
$R$-parity.  In this case, the lightest superpartner will provide
a second contribution to the dark matter.  Such multi-component DM
scenarios have been considered in a number of recent
works~\cite{Hur:2007ur,Feng:2008ya,Fairbairn:2008fb,Zurek:2008qg}.
 As long as any additional DM component has a small scattering
cross section off nuclei, it will not ruin the IDM explanation for
the DAMA signal provided the IDM component makes up a significant
fraction of the DM relic abundance.

\subsection{Models Mediated by an Exotic $Z'$ Gauge Boson}

  In Section~\ref{general} we showed that the DAMA signal can arise
from the scattering of IDM off iodine nuclei mediated by an exotic
$U(1)_x$ gauge boson.  We distinguish two possibilities for such a gauge
boson, one in which it is heavier than the SM $Z^0$ and couples directly to
the SM and the other in which it is  much lighter than the $Z^0$ and couples only
weakly to the visible sector, such as through a small kinetic mixing
with electromagnetism.  We construct here models for IDM that
realize both possibilities, though we will see the latter involves mass scales
that are more contrived.

\subsubsection{Heavy $U(1)_x$ Models}

  The $Z^0$-mediated candidates for IDM presented above
can all be adapted to models in which the scattering off nuclei
is mediated by the exchange of a heavy $U(1)_x$ gauge boson.
In each case, the $SU(2)_L$ doublet states in our previous models
are replaced by a pair of states with vector-like charges
under a $U(1)_x$ gauge symmetry.  For the analog of our SUSY model,
the $U(1)_x$ gauge symmetry should be broken by a pair of
fields such that they obtain hierarchically different expectation
values from the dynamics of the potential.

  In contrast to the $Z^0$-mediated models presented above, however,
the thermal relic density in heavy $U(1)_x$ models is more model dependent.
If the DM particle is lighter than the exotic gauge boson, it will no
longer annihilate into gauge boson pairs.  Annihilation
into matter fields from $s$-channel exchange of heavy gauge bosons
is suppressed by the larger gauge boson mass (and depends on the
number of channels into which the gauge boson can decay).
This can allow for  thermal dark matter that is somewhat lighter than in the
doublet case, which can provide a better IDM fit to the direct
detection data.

\subsubsection{A Light $U(1)_x$ Model}

  IDM scattering at DAMA can also be mediated by a new light  gauge
boson that couples directly to the DM, but only very weakly to the
visible SM sector.  This can arise from an exotic $U(1)_x$ gauge
boson with a mass on the order of a few GeV that has a kinetic
mixing with electromagnetism on the order of $\epsilon \sim
10^{-4}\!-\!10^{-3}$.  We describe a minimal supersymmetric
realization of such a scenario in this section.  However, in this
scenario and others like it, new mass scales must be put in by
hand.   Thus, we find this scenario less compelling from the point
of view of naturalness (not technical naturalness)
but permissible so we include it as a logical possibility.

  Supersymmetry affords a natural setting for light exotic gauge
bosons in sectors that are somewhat shielded from the source of
supersymmetry breaking. The smaller supersymmetry breaking soft
terms in the hidden sector can then induce symmetry breaking in that
sector at a scale that is parametrically smaller than the
electroweak scale.  However, the minimal model we describe below
also contains supersymmetric mass scales whose origin requires
further explanation.

  Our IDM model contains two pairs of SM-singlet
chiral superfields, $a$ and $a^c$ along with $H$ and $H^c$,
such that each pair has vector-like charges under a new $U(1)_x$
gauge symmetry.  The model also contains a pure singlet, $S$,
uncharged under both the SM and $U(1)_x$.
Both $a$ and $a^c$ as well as $S$ are assumed to be odd under
an exact unbroken $\mathbb{Z}_2$ symmetry.
We take the superpotential to be
\beq
W \supset \mu'\,H\,H^c + M_a\,a\,a^c + \frac{1}{2}\,M_s\,S^2
+ \lambda_1\,S\,a^c\,H + \lambda_2\,S\,a\,H^c,
\label{wlight}
\eeq
where we assume $M_a \sim M_s \sim \tev$ and $\mu'\sim \gev$.
We assume further that the fields in this sector are shielded from
supersymmetry breaking relative to the MSSM sector.  This can
hold if supersymmetry breaking is mediated to the visible sector
by gauge mediation through messengers charged only under the MSSM
gauge group, for example.

  In the context of gauge mediation, kinetic mixing with hypercharge
then induces effective charges for the gauge messengers under the
$U(1)_x$ symmetry that are suppressed by the mixing parameter.
The resulting soft scalar masses in the $U(1)_x$ sector are thus on
the order of $m_x^2 \sim (g_x/g')^2\,\epsilon^2\,m_{E^c}^2$, where
$\epsilon \sim 10^{-4}\!-\!10^{-3}$ is the kinetic mixing
parameter and $m_{E^c}^2$ is the soft scalar mass of the
right-handed selectron~\cite{Suematsu:2006wh,Hooper:2008im,
Zurek:2008qg,Chun:2008by}.
Upon running down to lower energies, the scalar soft masses for $H$
and $H^c$ can be induced to run negative by way of large Yukawa
couplings, generating VEVs for these fields on the order of a
GeV~\cite{Zurek:2008qg,Chun:2008by}. The $H$ and $H^c$ scalars can also be
destabilized at the origin by the contribution to the $U(1)_x$
$D$-term potential from the MSSM Higgs fields, which obtain small
$U(1)_x$ charges from gauge kinetic mixing.  The hidden sector
VEVs generated in this way will be on the order of
$\sqrt{\epsilon}\,v$.
If the scale of gauge mediation is relatively high, the $U(1)_x$
sector can also receive additional small soft breaking contributions
from gravity mediation~\cite{ArkaniHamed:2008qn,ArkaniHamed:2008qp}.

   Integrating out $S$ generates the effective superpotential
\beq
W_{eff} \supset \mu'\,H\,H^c + M_a\,a\,a^c
-\frac{\lambda^2_1}{2M_s}\,(a^cH)^2
-\frac{\lambda_2^2}{2M_s}\,(aH^c)^2 -
\frac{\lambda_1\lambda_2}{M_s}(a^cH)(aH^c). \eeq From this we
obtain the fermion mass splitting \beq \delta =
\frac{\lambda_1^2\left<H\right>^2+\lambda_2^2\left<H^c\right>^2}{M_s}
= 2\lambda^2\,\lrf{\left<H\right>}{\gev}^2\lrf{\tev}{M_s}\,\mev,
\eeq where in the second equality above we have assumed $\lambda_1
= \lambda_2 = \lambda$ and $\left<H\right> = \left<H^c\right>$.
With a small amount of suppression from the couplings, this is of
the right size for IDM.  From the $F$ terms of $H$ and $H^c$ we
also get scalar splittings \bea V_F &\supset&
-\mu'\frac{\lambda_1^2}{2M_s}\,H^{\dagger}H^c\,\tilde{a}^2
-\mu'\frac{\lambda_2^2}{2M_s}\,{H^c}^{\dagger}H\,{\tilde{a}^c}{}^2
\\
&&~~~ -
\frac{\lambda_1^2}{M_s}\,M_a\,(H^c)^2\,\tilde{a}\,{\tilde{a}^c}{}^*
- \frac{\lambda_2^2}{M_s}\,M_a\,(H)^2\,\tilde{a}^c\,{\tilde{a}}^*
+ h.c.. \nnmb
\eeq
The first two terms here are subleading, while the second two
generate scalar mass splittings of the right size with a small
amount of additional suppression from the couplings $\lambda_{1,2}$.
Note that either the fermion or the scalar can be the dark matter state,
whichever is lighter, depending on the soft terms.

   While the superpotential of Eq.~\eqref{wlight} is technically
natural, this model does not give an explanation for why $M_a$ and
$M_s$ are so much larger than $\mu'$.  A value of $\mu'$ on the
order of a GeV can perhaps arise naturally from an NMSSM-like
extension of the Higgs sector in this model, as in
Refs.~\cite{Zurek:2008qg,Chun:2008by}. It could also arise from a
Giudice-Masiero~\cite{Giudice:1988yz} coupling to supergravity in
high-scale gauge mediation~\cite{ArkaniHamed:2008qp}. The larger
masses $M_a$ and $M_s$ could also potentially be generated by an
NMSSM-like extension coupling to $a^{(c)}$ and
$S$~\cite{Chun:2008by}, although this possibility would be more
complicated.

  The model described above is similar to the one presented in
Refs.~\cite{ArkaniHamed:2008qn,ArkaniHamed:2008qp}.  There,
the larger mass is also put in by hand, the smaller mass is related
to the breakdown of a non-Abelian gauge symmetry near a GeV,
and the inelastic mass splitting arises radiatively from
gauge boson loops.
However, because this mechanism for inelastic splitting
works only with a non-Abelian gauge group,
the construction also requires a higher-dimensional operator
in order to generate kinetic mixing with hypercharge,
$Tr({\mathcal{O}^{(n)\,a}}\,W^a_{\mu\nu})\,B^{\mu\nu}/M^n$, where
${\mathcal{O}^{(n)\,a}}$ is a chiral adjoint operator under the
exotic gauge group of dimension $n$.
It requires additional model ingredients to explain the origin
of the scale $M$ which cannot be too large in order
to obtain a large-enough mixing angle.

\section{Conclusions\label{concl}}

  In this paper we have considered the possibility that DAMA
is a true discovery of dark matter, and investigated the
properties a theory of dark matter needs to have in order to
account for the data.  We have shown that inelastic dark matter is
consistent with the findings of DAMA as well as other direct
detection experiments and seems to provide a better account for
the sum of this data than other proposed explanations such as
light elastic dark matter or dark matter scattering off detector
electrons.  Extending the study of Ref.~\cite{Chang:2008gd}, we
find that heavier inelastic dark matter can give a reasonable fit
to the data, particularly for lower values of the galactic DM
escape velocity.

  An intriguing additional observation is that if the inelastic dark
matter candidate is an electroweak doublet, it can simultaneously
have the correct thermal relic abundance and a nucleon scattering
cross section mediated by the SM $Z^0$ in the range consistent
with DAMA and other experiments.  For a scalar doublet, this
occurs when its mass is close to $525\,\gev$, while for a fermion
doublet the mass should be about $1080\,\gev$.  This makes
electroweak-doublet inelastic dark matter candidates particularly
attractive.

The DAMA signal can also be explained by the inelastic scattering
of DM off nuclei mediated by a new massive $U(1)_x$ gauge boson.
This exotic gauge boson can either be heavy and couple directly to
the SM, or very light and hidden.  In the heavy case, the new
gauge boson must be somewhat leptophobic and have large couplings
to quarks and the DM.  Light gauge bosons can work if they have a
GeV-range mass and couple to the SM through a small kinetic mixing
with the photon.  In contrast to the case of electroweak doublet
DM, however, the correct nucleon scattering cross section and
thermal relic density do not arise automatically, and must be
arranged by hand.

  Given that inelastic dark matter gives a compelling explanation for
the DAMA result, it is of interest to understand what kind of models
of IDM might work.  The properties we need are clear.  We need a
particle of mass $\sim 100\!-\!1000\gev$ whose real and imaginary
(or Weyl) components are split by about $\sim 100\kev$. Such a small
splitting violates a $U(1)_{DM}$ global symmetry that is preserved
by a Dirac or complex scalar mass term. This makes such candidates
technically natural.

  We then address the question of overall naturalness. That is,
why should the mass splitting be six orders of magnitude smaller
than the overall mass scale?  We found several candidate models.
The first two work in the context of warped extra dimensions,
where $U(1)_{DM}$ symmetry breaking can be sequestered.  The
symmetry breaking resides either in a large fermion Majorana mass
or a small scalar holomorphic mass for a bulk singlet, that then
mixes with an $SU(2)_L$ doublet dark matter candidate.  The
downside of these models is that we do not yet know if a warped
extra dimension exists, or if the low UV scale needed in the
scalar model is present.

  The remaining models work in the context of supersymmetry.
In the most compelling supersymmetric model, the smallness of the
splitting arises through the mixing between an $SU(2)_L$ doublet
fermion dark matter candidate and a singlet which directly couples
to $U(1)_{DM}$ breaking.  The smallness of the mixing is
attributable in part to large $\tan\beta$.  The downside is that
we do not know if weak-scale supersymmetry is present, or why
there should be a minor conspiracy of small numbers to give the
necessary suppression.  Other supersymmetric models can work, but
they generally require more assumptions or more complications to
sufficiently isolate $U(1)_{DM}$ symmetry breaking.

  We do not yet know if the DAMA signal will prove to be new physics.
If it is, we conclude that inelastic dark matter particles are
excellent candidates.  Reasonable (but not completely obvious)
assumptions then lead to acceptable models of inelastic dark
matter.  We are very fortunate in that upcoming results from
XENON, CRESST, and other direct detection experiments should be
able to help determine if such models are likely correct.

\section*{Acknowledgements}

We thank Aaron Pierce, David Tucker-Smith, Doug Finkbeiner, John
Mason, John Ng, Kathryn Zurek, Matthew Schwartz, Neal Weiner, Nima Arkani-Hamed,
Paolo Gondolo, Pierluigi Belli, and Tracy Slatyer for helpful
comments and discussions. This work is supported in part by the
Harvard Center for the Fundamental Laws of Nature.  L.R. and D.P.
are supported by NSF grant PHY-0556111.


\appendix

\section{Appendix: Decoupling of KK Modes in the
Warped Fermion Model\label{appa}}

In this appendix we show that the effect of the bulk singlet KK
modes can be neglected when calculating the DM mass splitting in
the warped fermion model of section~\ref{seesaw}.  Note that this
is despite the fact that the zero mode and KK modes have a large
mixing due to the Majorana mass terms (\emph{i.e.}, $A_{0 0} \sim
A_{m n}$).  The key point is that the KK modes all have
vector-like $U(1)_{DM}$ preserving masses, and are very
inefficient at communicating $U(1)_{DM}$ breaking.

One can see this explicitly in the case of the first KK mode by
including the couplings to the dark matter doublet as part of the
mass matrix, and solving for the eigenvalues in the limit of a
small Higgs VEV.  For simplicity we can also neglect the Dirac
mass of the doublet, which plays no role in the communication of
$U(1)_{DM}$ breaking.  The mass matrix is then
\beq
\mathscr{L}
\supset - \frac12 \left(
\begin{array}{cccc} \bar{S}^0_L & \bar{S}^1_L & \bar{S}^{1 c}_R & \bar{D}^c_R
\end{array} \right) \left( \begin{array}{cccc}
  A_{0 0} & A_{0 1} & 0 & C_0 v \\
  A_{0 1} & A_{1 1} & m_1 & C_1 v \\
  0 & m_1 & 0 & 0 \\
  C_0 v & C_1 v & 0 & 0
\end{array}
\right) \left( \begin{array}{c} S_L^{0 c} \\ S_L^{1 c} \\ S_R^{1}
\\ D_R \end{array} \right) + h.c.,
\eeq
and the eigenvalues are determined from the roots of the
characteristic polynomial $\text{Det}(M - \lambda I)$
\bea
0 &=& \lambda^4 - (A_{0 0} + A_{1 1}) \lambda^3
- (C_0^2 v^2 + C_1^2 v^2 + m_1^2) \lambda^2 \\
    &&  + \left( A_{0 0} (m_1^2 + C_1^2 v^2)
+ A_{1 1} C_0^2 v^2 - 2 A_{0 1} C_0 C_1 v^2 \right) \lambda
+ m_1^2 C_0^2 v^2
\nonumber,
\eea
where we have used the fact that $A_{0 0} A_{1 1} - A_{0 1}^2
= 0$.  In the limit that $v^2$ goes to 0, there is a zero
eigenvalue corresponding to the doublet.  Thus we expect that the
eigenvalue is proportional to $v^2$.  Plugging in $\lambda = x
v^2$ and dropping terms of $O(v^4)$, we obtain
\beq
0 \approx x
A_{0 0} m_1^2 v^2 + m_1^2 C_0^2 v^2 + O(v^4)
\eeq
so $x \approx \frac{- C_0^2}{A_{0 0}}$, and we see that at leading order the
Majorana mass of the doublet is $\lambda \approx \frac{- C_0^2
v^2}{A_{0 0}}$.  This only depends on $A_{0 0}$, and is
independent of $A_{1 1}$ and $A_{0 1}$ as we assumed in the
estimates of section~\ref{seesaw}.

The leading corrections to this formula are proportional to
$\sim \frac{C^4 v^4}{A m^2}$ and are
suppressed compared to the contribution from the zero mode by
a factor $\sim \frac{v^2}{m^2}$.  We have numerically checked
that the effects of including more KK modes are also similarly
 suppressed, and that the sequence rapidly converges after the
first few modes.  We also note that this behavior is consistent
with the results of Ref.~\cite{Huber:2003sf}.

\section{Appendix: Decoupling of Higher KK Modes in the
Warped Scalar Model\label{appb}}

  In this appendix we will show that the effective coupling of
singlet KK modes to the DM doublet $C_n^{eff}$ becomes suppressed
as $\sim 1/n$ for large $n$ in the warped scalar model of
section~\ref{scalarmodel}.

 Substituting the form of the KK mode wavefunctions
Eq.~\eqref{scalarwavefunction} into Eq.~\eqref{cneffeq}, we see
that computing $C_n^{eff}$ requires performing the integral
\beq
C_n^{eff} \simeq \sqrt{\frac{\pi m_n}{2}} \frac{\lambda e^{-\pi k
R/2}}{\Delta} \int_{\pi R - \frac{\Delta}{2}}^{\pi R +
\frac{\Delta}{2}} d y \,\left[ J_{\alpha}(\frac{m_n}{k} e^{k y}) +
b_{\alpha}(m_n) Y_{\alpha}(\frac{m_n}{k} e^{k y})\right].
\eeq
Close to $y = \pi R$, the integrand is completely dominated by
$J_{\alpha}$.  For large argument (large $n$), this Bessel function
can be approximated as
\beq
J_{\alpha}(\frac{m_n}{k} e^{k y})
\approx \sqrt{\frac{2 k}{\pi m_n e^{k y}}} \cos\left(\frac{m_n}{k}
e^{k y} - \frac{\pi}{2}(\alpha + \frac12)\right).
\eeq
Changing variables to $z = \frac{1}{k} e^{k y}$ then gives
\beq
C_n^{eff}
\simeq \frac{\lambda e^{-\pi k R/2}}{k \Delta}
\int_{z_{-}}^{z_{+}} d z\, z^{- \frac32} \,\cos\left(m_n z -
\frac{\pi}{2}(\alpha + \frac12)\right),
\eeq
where $z_{\pm} = \frac{1}{k} e^{k (\pi R \pm \frac{\Delta}{2})}$. The cosine
rapidly oscillates over the brane thickness at large n, while the
$z^{- \frac32}$ piece is relatively stable. Approximating the
stable piece by its central value and performing the integral, one
obtains
\beq
C_n^{eff} \approx \frac{\lambda \sqrt{k} e^{-2 \pi k
R}}{m_n \Delta} \left[\sin\left(m_n z_{+} - \frac{\pi}{2}(\alpha +
\frac12)\right) - \sin\left(m_n z_{-} - \frac{\pi}{2}(\alpha +
\frac12)\right)\right],
\eeq
and then one can place an approximate upper bound the magnitude
of $C_n^{eff}$,
\beq |C_n^{eff}| \lesssim
\frac{\lambda \sqrt{k} e^{-2 \pi k R}}{m_n \Delta}.
\eeq
Thus, we see that the magnitude of $C_n^{eff}$ falls roughly as
$\sim 1/n$ as we set out to show.



\end{document}